\documentclass[12pt,aps]{revtex4}
\usepackage{epsfig,amsmath,amssymb}

\newcommand{\eq}{{ \equiv }}
\newcommand{\fr}[1]{
             \frac{#1}}
\newcommand{\bea}{\begin{eqnarray}}
\newcommand{\eea}{\end{eqnarray}}

\newcommand{\chibar}{\overline{\chi}}

\newcommand{\bbar}{B\,-\,\overline{B}}
\newcommand{\Qp}{Q^{(+)}_v}
\newcommand{\Qm}{Q^{(-)}_v}
\newcommand{\ket}{{\rangle}}

\newcommand{\bra}{{\langle}}
\newcommand{\gc}{\bra\fr{\alpha_s}{\pi}G^2\ket}

\newcommand{\qc}{\bra\,\overline{q}q\,\ket}
\newcommand{\ga}{g_{{\cal A}}}

\def\slash#1{#1 \hskip -0.5em / }

\begin{document}
\title{Chiral quark models and their applications
\footnote{Dedicated to the memory of Prof. Dubravko Tadi\'c}}

\author{Jan~O.~Eeg}
\email{j.o.eeg@fys.uio.no}
\affiliation{Department of Physics, University of Oslo,
P.O.Box 1048 Blindern, N-0316 Oslo, Norway}

\author{Aksel~Hiorth}
\email{aksel.hiorth@rtf.no}
\affiliation{RF-Rogaland Research, P.O.Box 8046, N-4068 Stavanger, Norway}
\begin{abstract}
We give an overview of chiral quark models, both for the pure light
sector and the heavy-light sector.
 We describe how such models can be bosonized to obtain
well known   chiral Lagrangians which can be inferred
 from the symmetries of QCD alone. In addition, we can within these
 models  calculate the coefficients of the various pieces of the chiral Lagrangians.
We discuss a few applications of the models,
 in particular,   $\bbar$ mixing and processes of the type  $B \rightarrow D \bar{D}$,
where $D$ might be both pseudoscalar and vector. We suggest
how the formalism might be extended to include light vectors
($\rho,\omega,K^*$), and heavy to light transitions like $B \to \pi$. 

\end{abstract}
\maketitle

\section{Introduction}

While the short distance (SD) effects in hadronic  physics are well
 understood within perturbative quantum chromodynamics (pQCD),
 long distance (LD) effects
 have been hard to pin down.
Even if quark models are not QCD itself,
various  QCD inspired quark models have 
 been useful to make predictions for a
limited class of problems. Lattice QCD and QCD sum rules are 
on more solid ground theoretically, but are in various cases not so
easy to apply.
In the light quark sector,
low energy quantities have been studied in terms of the (extended)
 Nambu-Jona-Lasinio model (NJL)\cite{bijnes},
and  also  the chiral quark model ($\chi$QM)\cite{chiqm}, which is the mean
 field approximation of NJL. 

\begin{figure}[t]
\begin{center}
\epsfig{file=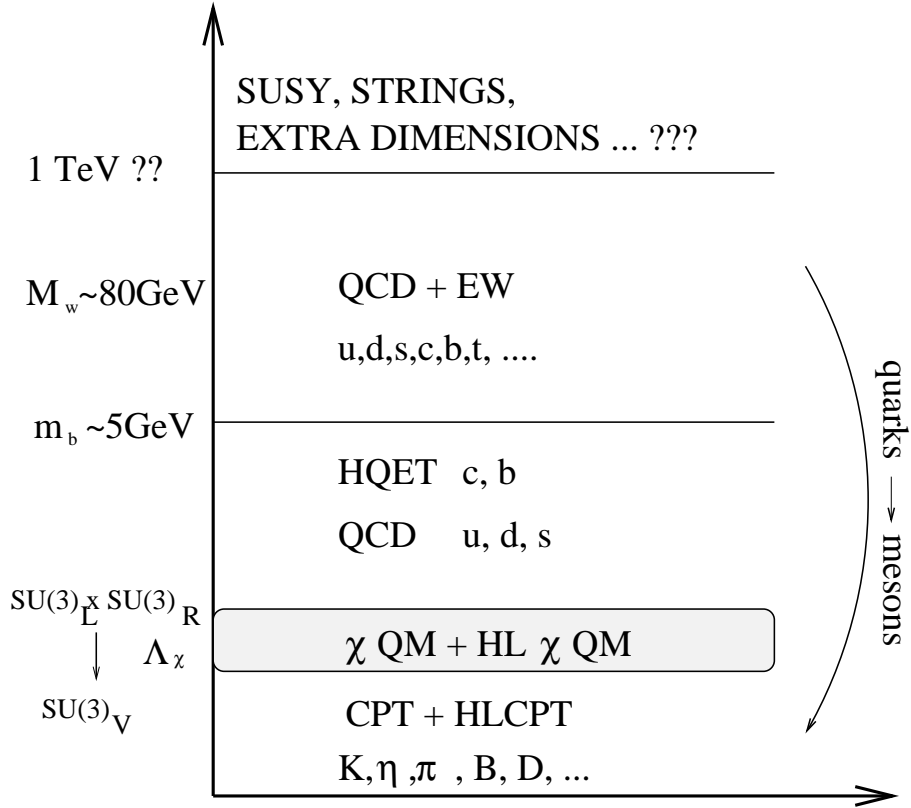,width=12cm}
\caption{Energy hierarchy in interactions of elementary particles.}
\label{fig:energylev}
\end{center}
\end{figure}

Within the $\chi$QM, 
the light quarks ($u, d, s$) couple to the would be Goldstone octet
mesons ($K, \pi, \eta$) in a chiral invariant way, such that all effects 
are in principle calculable in terms of physical quantities and a few
model dependent parameters, namely the quark condensate, the gluon condensate,
and the constituent quark mass \cite{pider,epb,BEF}. More specific, one 
may calculate the coupling constants of chiral Lagrangians
by integrating out the quarks by means of the $\chi$QM.
In this way
chiral quark models bridge between  pQCD and chiral perturbation
theory ($\chi PT$) as indicated in Fig.~\ref{fig:energylev}.

The ideas from the chiral quark model
of the pure light sector \cite{chiqm,pider,epb,BEF} has  been extended to
 the sector 
involving a heavy quark ($c$ or $b$) and thereby to  heavy-light
mesons \cite{HLchiqm}. Such models we name heavy-light chiral quark 
models (HL$\chi$QM).
 Also in this case, one may integrate out
the light and heavy quarks and  obtain chiral Lagrangians involving
light and heavy mesons \cite{wisitchpt}. That is, we
 calculate the parameters
 of chiral Lagrangian terms,
where the description of  heavy mesons are in accordance with heavy quark 
effective field theory (HQEFT) \cite{neu}.
In our approach \cite{AHJOE} we extended the formalism of
\cite{HLchiqm}  to include gluon (vacuum) condensates.

 One important 
motivation for the inclusion of gluon condensates is
the possibillity to estimate non-factorizable (colour suppressed) 
contributions in non-leptonic
decays. For instance, 
$K- \bar{K}$-mixing and 
 the $\Delta I  =   1/2$ rule for 
$K \rightarrow 2 \pi$ can be understood in a reasonable way
within the $\chi$QM \cite{pider,BEF} including gluon condensates.
Especially, the suppression of the $I=2$ amplitude found for $K \rightarrow 2 \pi$
 is also in agreement with generalized 
factorization \cite{cheng}.
 Furthermore, it allows us for instance to consider decays where
 the gluonic aspect of $\eta'$ is relevant \cite{EHP}, and
some aspects of $D$-meson decays \cite{EFZ}.
The most important application is to 
 calculate non-factorizable 
 contributions to $B-\bar{B}$-mixing \cite{ahjoeB}, where our approach
 includes $1/m_b$ corrections and chiral corrections both from loops
 and counterterms.
Also processes of the type  $B \rightarrow D \bar{D}$ are
 calculable \cite{EFH}.
 It should  be emphasized that the HL$\chi$QM 
can not, -in its present form,  be used for heavy to light
 transitions like $B \rightarrow  \pi K$, where QCD factorization
 \cite{BBNS} or soft collinear theory(SCET) \cite{SCET} is often applied .
Still, in the last section, we suggest how an extension to this case
might be performed. We also suggest how the $\chi$QM might be extended
to include light vectors ($\rho,\omega,K^*$).

\section{Chiral perturbation theory}

\subsection{The pure light sector}

Quarks are the fundamental hadronic matter. However, the particles we observe
 are those built out of them: baryons and mesons. In the sector 
of the lowest
mass pseudoscalar mesons (the would-be Goldstone bosons: $\pi$,
$K$ and $\eta$) the interactions can be described in terms of an effective
theory, the chiral Lagrangian, that includes only these states.
The chiral Lagrangian and chiral perturbation 
theory ($\chi$PT)~\cite{Weinb,GassL}
provide  a faithful representation of 
this sector of the Standard Model  after the
 quark and
gluon degrees of freedom have been integrated out.  The form of
this effective field theory and all its possible terms
are determined by $SU_L(3) \times SU_R(3)$
chiral invariance and Lorentz invariance.
Terms which explicitly break chiral invariance are introduced in terms
of the quark mass matrix ${\cal M}_q$. 

The strong chiral lagrangian is completely fixed to the leading order in
momenta by symmetry requirements and the Goldstone boson's decay
amplitudes:
\bea
{\cal L}_{\rm strong}^{(2)} = 
\frac{f^2}{4} Tr \left( D_\mu \Sigma D^\mu \Sigma^{\dag} \right)
 +  \frac{f^2}{2}
B_0 Tr \left( {\cal M}_q \Sigma^{\dag} +  \Sigma {\cal M}_q^{\dag} \right) \, ,
\label{L2strong}
\eea
where the covariant derivative $D^\mu$ contains the photon field,
and  ${\cal M}_q = \mbox{diag} [ m_u, m_d, m_s ]$. The quantity  $B_0$ is defined by
$\langle \bar{q}_i q_j \rangle = - f^2 B_0 \delta_{ij}$, where
\bea
(m_s + m_d) \langle \bar{q} q \rangle   =  - f_K^2 m_K^2  \; , \quad
(m_u + m_d) \langle \bar{q} q \rangle   =  - f_\pi^2 m_\pi^2  \; ,
\label{PCACon}
\eea
in the PCAC limit. The quantity $\langle \bar{q} q \rangle$ is the
quark condensate, being of  order~(-240 MeV$)^3$.
The $SU_L(3) \times SU_R(3)$ field $\Sigma$ contains the pseudoscalar octet  $\Pi$:
\bea
\Sigma \equiv \exp \left( \frac{2i}{f} \,\Pi   \right)
\; \, ,
\quad 
 \Pi=\fr{\lambda^a}{2}\phi^a(x) = 
\frac{1}{\sqrt{2}} \left[\begin{array}{ccc} \fr{\pi^0}{\sqrt{2}}
+\fr{\eta_8}{\sqrt{6}} & \pi^+
&K^+\\ \pi^-&-\fr{\pi^0}{\sqrt{2}}+\fr{\eta_8}{\sqrt{6}} & K^0\\
K^- &\overline{K^0}& -\fr{2}{\sqrt{6}}\eta_8\end{array}\right] \; .
\label{sigma}
\eea
The quantity $f$ is, to lowest order,  identified with the  pion decay constant $f_\pi$
(and equal to $f_K$ before chiral loops are introduced).

 When the
matrix $\Sigma$ is expanded in powers of $f^{-1}$, the zeroth
 order term obtained from (\ref{L2strong}) is the 
free  Klein-Gordon Lagrangian for the pseudoscalar particles.
From this Lagrangian one might deduce the (left-handed) current
\bea
J^n_\mu = -i \frac{f^2}{2}  Tr \left[ \lambda^n \Sigma D^\mu \Sigma^{\dag}
  \right] \; ,
\label{Lightcurr}
\eea
where $n$ is a flavour octet index and $\lambda_n$ a $SU(3)$ flavour matrix.

For  the next-to-leading order Lagrangian ${\cal L}_{\rm strong}^{(4)}$
 there are ten terms and thereby
ten coefficients $L_i$ to be 
determined~\cite{GassL} either 
experimentally or by means of some
model.  Some of these play an important role in the physics 
of $\epsilon'$ in  $K \rightarrow 2 \pi$ decays\cite{epsK}.
 As  examples, we display the $L_5$ and $L_8$ terms in
 governing much of the
penguin physics:
\bea
L_5 \, B_0 \, Tr \left[ D_\mu \Sigma^{\dag} D^\mu \Sigma
 \left( {\cal M}_q^{\dag} \Sigma +  \Sigma^{\dag} {\cal M}_q \right)
 \right] \; ,
\label{L5strong}
\eea
and
\bea
L_8 \, B_0 \, Tr \left[ {\cal M}_q^{\dag} \Sigma  {\cal M}_q^{\dag} \Sigma +
 {\cal M}_q \Sigma^{\dag}  {\cal M}_q \Sigma^{\dag} \right] \, .
\label{L8strong}
\eea

Under the action
of the elements $V_R$ and $V_L$ of the chiral
group $SU_R(3) \times SU_L(3)$, the field $\Sigma$ transforms as:
\bea
\Sigma \rightarrow  V_L \Sigma V_R^{\dag} \; ,
\label{Sitransf}
\eea
and accordingly for the conjugated fields.
Formally, ${\cal M}_q$ is given the same transformation properties as
$\Sigma$,  and ${\cal M}_q^\dagger$ as  $\Sigma^\dagger$.

\subsection{The heavy light sector}

The strong chiral Lagrangian for the heavy light sector is 
 \cite{wisitchpt,HLchpt}:
\bea
{\cal L}_{Str}\, &&=   \, \mp  
Tr\left[\overline{H^{(\pm)}_{v k}}(iv\cdot {\cal D}_{hk})H^{(\pm)}_{v h}\right]  \,
- \, g_{\cal A}  
Tr\left[\overline{H^{(\pm)}_{v k}} H^{(\pm)}_{v h}\gamma_\mu\gamma_5 {\cal
A}^\mu_{hk} \right] \nonumber \\ 
&&+ 2 \lambda_1 Tr\left[\overline{H^{(\pm)}_{v k}} H^{(\pm)}_{v h} 
(\widetilde{M}_q^V)_{hk}\right]
+ \frac{e \beta}{4}Tr[\overline{H^{(\pm)}_{v k}} H^{(\pm)}_{v h} \sigma\cdot
F \, (Q_q^\xi)_{hk}]\, + ......
\label{LS0}
\eea
where $k,h$ are $SU(3)$ triplet indices, and $v$ is the velocity of the heavy meson. 
The ellipses indicate other terms (of higher order, say), 
and $i {\cal D}_{hk}^\mu = \delta_{hk} D^\mu +  {\cal V}^\mu_{hk}$.
Moreover, $Q^\xi_q=(\xi^\dagger Q_q \xi+\xi Q_q \xi^\dagger)/2$, where $Q_q$ is the
$SU(3)$ charge matrix for light quarks, $Q_q = diag(2/3,-1/3,-1/3)$,
 and $F$ is the electromagnetic field tensor.
$H_{v k}^{(\pm)}$ is the heavy meson field  containing
 a spin zero and spin one boson:
\begin{eqnarray}
&H_{v k}^{(\pm)} & \eq  P_{\pm}(v) (P_{k}^{(\pm) \mu} \gamma_\mu -     
i P_{k}^{(\pm) 5} \gamma_5)\; , \nonumber \\
&\overline{H_{v k}^{(\pm)}}
& =  \gamma^0 (H_{v k}^{(\pm)})^\dagger \gamma^0
 =  \left[(P_{k}^{(\pm)\mu})^{\dagger} \gamma_\mu 
 -   i (P_{k}^{(\pm) 5})^\dagger \gamma_5\right] P_{\pm} \; , \label{barH}
\end{eqnarray}
where
\begin{equation}
P_{\pm}(v) =  (1 \pm \gamma \cdot v)/2 
\label{proj}
\end{equation}
are projection operators. 
The  fields $P^{(\pm) 5} (P^{(\pm)\mu})$ represent  heavy-light mesons,
 $0^{-}(1^-)$,  with velocity $v$. The signs $\pm$ refers to particles
 and anti-particles respectively, and will  sometimes be omitted in the
 following when unnecessary.

The vector and axial vector fields 
${\cal V}_{\mu}$ and  
${\cal A}_\mu$ are given by:
\begin{equation}
{\cal V}_{\mu}\eq \fr{i}{2}(\xi^\dagger\partial_\mu\xi
+\xi\partial_\mu\xi^\dagger 
) \qquad ;  \qquad  
{\cal A}_\mu\eq  -  \fr{i}{2}
(\xi^\dagger\partial_\mu\xi
-\xi\partial_\mu\xi^\dagger) \; .
\label{defVA}
\end{equation}

The fields $\xi$  and $H_v$  transform  as
\begin{equation}
 \xi  \rightarrow U  \, \xi  \, V_R^\dagger \,  =    V_L \xi U^\dagger
 \quad , \qquad  H_{v} \rightarrow  H_{v} \, U^\dagger  \; ,
\label{xiHtransf}
\end{equation}
where $U \, \epsilon \; SU(3)_V$, the unbroken symmetry.

The vector and axial fields transform as 
\begin{equation}
{\cal V}_{\mu} \rightarrow U \, {\cal V}_{\mu} \, U^\dagger \;  +   \; 
i U \partial_\mu \, U^\dagger \quad , \qquad 
{\cal A}_\mu \rightarrow  U {\cal A}_\mu \, U^\dagger \; . 
\end{equation}
 The vector field
${\cal{V}}^\mu$ is seen to transform as a gauge field under local $SU(3)_V$,
 and can only appear in  combination with a derivative as
a covariant derivative $( i \partial^\mu  +   {\cal{V}}^\mu)$.
The quantity  $\widetilde{M_q}^V$ (as well as the orthogonal combination
 $\widetilde{M_q}^A$) is related to the current  mass term:
\bea
\widetilde{M}_q^V \, \eq \, 
\fr{1}{2}(\xi^\dagger {\cal M}_q \xi^\dagger \,
+ \xi {\cal M}_q^{\dag} \xi )\quad ; \qquad 
\widetilde{M}_q^A\eq -\fr{1}{2}(\xi^\dagger {\cal M}_q \xi^\dagger
 -\xi {\cal M}_q^{\dag} \xi) \; .
\label{masst}
\eea

The heavy-light weak current, to zeroth order in $1/m_Q$ and chiral counting,
is represented by:
\begin{equation}
J_k^\alpha (0)   =    \fr{\alpha_H}{2} Tr\left[\xi^{\dagger}_{hk}\Gamma^\alpha
  H_{v h}\right]
  \; ,\label{J(0)}
\end{equation}
 and under $SU(3)_L$ it transforms as
\begin{equation}
J_k^\alpha \rightarrow J_h^\alpha \left( V_L^\dagger \right)_{h k} \; .
\label{Jtransf}
\end{equation}
This  current has  also 
 (counter) terms, of higher order in the chiral counting, 
 needed to make the chiral loops finite:
\begin{equation}
J^\mu_k({\cal M}) = 
\fr{\omega_1}{2}Tr[\xi^\dagger_{h k}\,\Gamma^\mu {H_v}_{l}\,
\widetilde{{\cal M}^V_{l h}}]
+\fr{\omega_1^\prime}{2}Tr[\xi^\dagger_{kh}\,\Gamma^\mu {H_v}_{h}]\,
\widetilde{{\cal M}^V_{ll}} \; ,
\label{J2M}
\end{equation}
where the parameters $\omega_1$ and $\omega_1'$ are commented on in
section IV-C. 
To leading order,
 $\Gamma^\alpha = \gamma^\alpha L$, where $L$ is the left -  handed
projector in Dirac space, 
$L =  (1 -  \gamma_5)/2$.  However, this is  slightly modified by
perturbative QCD for $\mu$ below $m_Q$, which gives
 \cite{neu}
\begin{eqnarray}
\Gamma^\alpha \,\eq\, C_\gamma (\mu )\,\gamma^\alpha\, L \,+\,  
C_v(\mu )\, v^\alpha\, R\; ,
\label{Gamma}
\end{eqnarray}
where  $R$ is the  right -  handed projector, $R =  (1 +  \gamma_5)/2$ .
The coefficients $C_{\gamma,v}(\mu)$ are determined 
by QCD renormalization for  $\mu < m_Q$. They have been calculated to
NLO and the result is the same in $MS$ and $\overline{MS}$ scheme\cite{Cgamma}.
($C_\gamma$ is close to one and $C_v$ is rather small).
Corrections to the weak current of order $1/m_Q$ will be discussed in 
section \ref{sec:mq}.

Before closing this section, we write down 
the  bosonized $b \rightarrow c$ transition current in  terms of the
heavy fields
\begin{equation}
 \overline{Q_{v_b}^{(+)}} \,\gamma^\mu  L Q_{v_c}^{(+)}\;\longrightarrow
 \; - \zeta(\omega) Tr\left[ \overline{H_c^{(+)}}\gamma^\alpha L
 H_{b}^{(+)} \right] , 
\label{Jbc}
\end{equation}
where $\zeta(\omega)$ is the Isgur-Wise function for the $\bar{B}
\rightarrow D$ transition \cite{isgur}.
The indices  on the heavy fields here refer to the the
 $b$- and $c$-quarks with velocities $v_b$
 and $v_c$, with $\omega \equiv v_b \cdot v_c$.  
The current for 
 $D \overline{D}$ production is :
\begin{equation}
 \overline{Q_{v_c}^{(+)}} \,\gamma^\mu\, L Q_{\bar{v}}^{(-)}\; 
 \longrightarrow \;
   - \zeta(-\lambda) Tr\left[
 \overline{H_c^{(+)}} \gamma^\alpha L  H_{\bar{c}}^{(-)} \right] \; ,
\label{Jcc}
\end{equation}
where  the Isgur-Wise function $\zeta(-\lambda)$ is (in general) complex.
We have $\lambda =v_c \cdot \bar{v} \, $, where $\bar{v}$ is the velocity
of $\bar{c}$.

\section{The chiral quark model ($\chi$QM)}

\subsection{The Lagrangians for $\chi$QM }\label{sec:CQM}
The light quark sector is described by the chiral quark model ($\chi$QM),
having a standard QCD term and a term describing interactions between
quarks and  (Goldstone) mesons \cite{bijnes,chiqm,epb,pider,BEF}: 
\begin{equation}
{\cal L}_{\chi QM} =   \bar{q}_L i \gamma \cdot D \, q_L \,  +   \, 
\bar{q}_R i \gamma \cdot D \, q_R
 -   \bar{q}_L {\cal M}_q \, q_R \,  -    \bar{q}_R {\cal M}_q^\dagger \, q_L
  -     m(\bar{q}_R \Sigma^{\dagger} q_L   +    \bar{q}_L \Sigma q_R) \; , 
\label{chqmu}
\end{equation}
where $m$ is the ($SU(3)$ -  invariant) constituent quark mass for light quarks
$q^T  =  (u,d,s)$. The left- and
 right-handed
 projections $q_L$ and $q_R$ are transforming after $SU(3)_L$ and $SU(3)_R$
respectively:
\bea
q_L  \rightarrow V_L \, q_L \qquad \mbox{and} \qquad q_R \rightarrow  
V_R \, q_R \; .
\label{qtransf}
\eea

From (\ref{chqmu}) we deduce the Feynman rules. For instance, the
$P q \bar{q}$ coupling is $(m \gamma_5/f)$ times some $SU(3)$ factor
($P$ is a pseudoscalar meson $\pi, K, \eta$). From
  such Feynman rules, and including the quark propagator
$S(p) = (\gamma \cdot p -M_q)^{-1}$, we can calculate aplitudes for, say, $\pi -\pi$
  scattering in the strong sector. Here $M_q= m+m_q$ is the total
  mass. Alternatively one might keep only the constituent mass $m$  in
  the propagator, and take the current mass $m_q$ as  a coupling. 
Incuding also the Feynman rules for weak vertices, one might
calculate \cite{BEF,epb}
amplitudes for non-leptonic decays in terms of quark loops
representing $f_\pi$ and the semileptonic form factors $f_\pm$, but
also for more complicated cases.

\begin{figure}[t]
\begin{center}
\epsfig{file=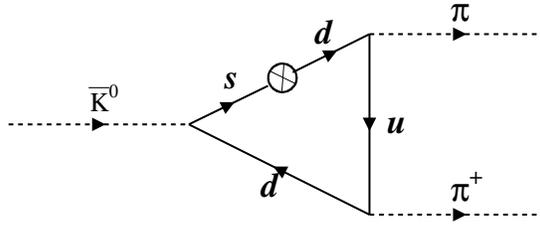,width=12cm}
\caption{Contribution to the process $K \rightarrow 2 \pi$
from the non-diagonal $s \rightarrow d$ transition.}
\label{fig:K2pi}
\end{center}
\end{figure}

Also, as a more exotic example, one may calculate the effect of the
electroweak $s \rightarrow d$ self-energy transition contribution to
$K \rightarrow 2 \pi$ as shown in Fig. \ref{fig:K2pi}. This is an
off-shell effect which vanish in the free quark case,
 but  is non-zero for bound quarks and 
proportional to  $m$  within our framework
\cite{epb}.

Chiral Lagrangians, either in the pure strong sector or for
non-leptonic decays, 
are however easier to obtain 
in a  more transparent way
within the  ``rotated version'' 
of the $\chi$QM 
with  flavour rotated quark fields $\chi$ given by:
\begin{equation}
\chi_L  =   \xi^\dagger q_L \quad ; \qquad \chi_R  =   \xi q_R \quad ; \qquad 
\xi \cdot \xi  =   \Sigma \; .
\label{rot}
\end{equation}
 The constituent quark fields 
$\chi_L$ and $\chi_R$ 
transform in a simple way under $SU(3)_V$:
\begin{equation}
\chi_L \rightarrow U \, \chi_L \quad , \qquad \chi_R \rightarrow U \; \chi_R
\; .
\label{chHtransf}
\end{equation}

In the rotated version, the chiral interactions are rotated  into the
kinetic term while the interaction term proportional to $m$ in 
(\ref{chqmu}) 
 become a pure (constituent) mass term \cite{chiqm,BEF}:
\begin{equation}
{\cal L}_{\chi QM} =  
\chibar \left[\gamma^\mu (i D_\mu   +    {\cal V}_{\mu}  +  
\gamma_5  {\cal A}_{\mu})    -    m \right]\chi 
  -     \chibar \widetilde{M_q} \chi \;  , 
\label{chqmR}
\end{equation}
which is manifestly invariant under $SU(3)_V$. Moreover,
\bea
  \widetilde{M_q} \eq \widetilde{M}_q^V   +    \widetilde{M}_q^A
  \gamma_5  \; ,
\label{cmass}
\eea
where $\widetilde{M}_q^{V,A}$ are given in (\ref{masst}).

In the light  sector, the various pieces of the strong Lagrangian
in section II-A
can now be obtained by integrating out the constituent quark fields $\chi$,
and these pieces can be written in terms of the  fields ${\cal A}_\mu \, ,
\, \widetilde{M}_q^V$ and $\widetilde{M}_q^A$
This can easily be seen by  using  the relation
\begin{equation}
{\cal{A}}_\mu \; =  \;  -   \fr{1}{2 i} \xi \, (D_\mu \Sigma^\dagger) \, \xi
\; = +\; \fr{1}{2 i} \xi^\dagger \, (D_\mu \Sigma) \, \xi^\dagger \; .
\label{ASigma}
\end{equation}

 In our model, the hard gluons are integrated out and we are
left with soft gluonic degrees of freedom. These gluons can be
described using the external field technique, and their
effect will be parameterized by vacuum expectation values, 
i.e. the gluon condensate $\gc$. Gluon condensates with higher 
dimension could also be included, but we truncate the expansion by keeping
 only the condensate with lowest dimension.

When calculating the soft gluon effects in terms of the gluon condensate,
we follow the prescription given in \cite{nov}.
\begin{figure}[t]
\begin{center}
\epsfbox{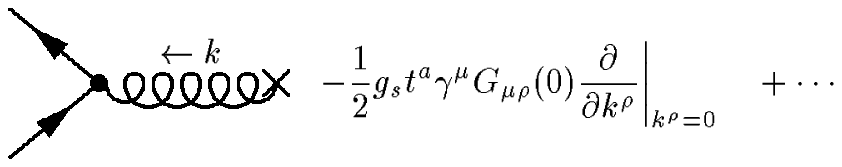}
\epsfysize=5cm
\caption{Feynmanrule for the light quark -soft gluon  vertex.}
\label{fig:gc}
\end{center}
\end{figure} 
The calculation  is easily carried out in the
Fock -  Schwinger gauge. In this gauge one can expand the gluon field as :
\begin{equation}
A_\mu^a(k) =   -  \fr{i(2\pi)^4}{2}G^a_{\rho\mu}(0)\fr{\partial}{\partial
k_\rho}\delta^{(4)}(k)  +  \cdots\, .
\end{equation}
In some simple cases one may also use  the light quark propagator in a
gluonic background (to first order in the gluon field):
\begin{equation}
S_1(p,G) = - \frac{g_s}{4} G^b_{\alpha \beta} t^b
\left[ \sigma^{\alpha \beta} (\gamma \cdot p + m) \, + \,
(\gamma \cdot p + m) \sigma^{\alpha \beta}\right]
(p^2-m^2)^{-2} \; ,
\label{S1G}
\end{equation}
where $g_s$ is the strong coupling constant, $a$ and $b$
are colour octet indices, and $t^a$ are the colour matrices.
In general one should stick to the
prescription in \cite{nov} in order to get correct results.
Since each vertex in a Feynman diagram is accomplished with an
integration we get the Feynman rule given in Fig.~\ref{fig:gc}.
The gluon condensate contributions are obtained by the replacement
\begin{equation}
g_s^2 G_{\mu \nu}^a G_{\alpha \beta}^b  \; \rightarrow \fr{4 \pi^2}{(N_c^2-1)}
\delta^{a b} \gc \frac{1}{12} (g_{\mu \alpha} g_{\nu \beta} -  
g_{\mu \beta} g_{\nu \alpha} ) \, .
\label{Gluecond}
\end{equation}.

\subsection{Bosonization of the $\chi$QM}

The Lagrangians \eqref{chqmu} or  \eqref{chqmR} from the previous section can now
be used for bosonization, i.e. to
integrate out the quark fields. This can be done in the path
integral formalism, or as we do here,
by expanding in terms of  Feynman diagrams.
Within the $\chi$QM, with Feynman rules obtained from (\ref{chqmu})
 one may calculate the simple quark loop amplitude
for $\pi \rightarrow W$ which defines  $f$ (the bare $f_\pi$)
in terms of  a logarithmicly
divergent integral $I_2$
times the coupling $\sim m/f$. Including also the  gluon condensate
contribution  one obtains
\cite{pider,epb,BEF}:
\begin{equation}
f^2 =    -  i4m^2N_cI_2 +  \fr{1}{24m^2}\gc \; ,
\label{fpidiv}
\end{equation}
where $I_2$ is the following logarithmic divergent integral ($d$ is
the dimension of space within dimensional regularization):
\bea
I_2\, &\eq\, &\int\fr{d^dk}{(2\pi)^d}\fr{1}{(k^2 -  m^2)^2} \, .
\label{I2int}
\eea

Equivalently, one may obtain the (kinetic part of the) strong Lagrangian in 
(\ref{L2strong}) by attaching two  axial fields  ${\cal A}_\mu$ to
a vacuum polarization like  quark loop diagram by using   \eqref{chqmR}. 
Then  one obtains:
\begin{equation}
 i {\cal{L}}^{(2)}_{str} \,
= - N_c \, \int \frac{d^dp}{(2\pi)^d} \,
Tr \, \left[ \left( \gamma_\sigma \gamma_5 {\cal{A}}^\sigma \right)
 \, S(p) \,
\left( \gamma_\rho \gamma_5 {\cal{A}}^\rho \right)  \,  S(p) \right]
 \, = \,
f^2 Tr \, \left[{\cal{A}}_\mu {\cal{A}}^\mu \right] \;,
\label{L2str}
\end{equation}
where  the trace is
both in flavor and Dirac spaces 
(a similar diagram with glouns should also be added).
This is easily seen by using the relation (\ref{ASigma}), provided 
$f^2 $ is given by (\ref{fpidiv}). The eq. (\ref{L2str}) give the 
 Lagrangian (\ref{L2strong}) in the light sector by applying (\ref{ASigma}).

The quark condensate is: 
\begin{equation}\label{I1}
\qc = -i N_c Tr \int\frac{d^dp}{(2\pi)^d} S(p) \, 
= \, -4imN_cI_1-\fr{1}{12m}\gc \; ,
\end{equation}
 where $I_1$ is the quadratically divergent integral
\bea
I_1 \, &\eq &\,\int\fr{d^dk}{(2\pi)^d}\fr{1}{k^2 -  m^2} \; . 
\label{I1int}
\eea
Here the propagator $S$ has to be understood at
the one in the gluon field up to second order
and  the a priori 
divergent integrals $I_{1,2}$ have to be interpreted as  the
regularized ones.
The physical values of $I_{1,2}$ are
determined  by 
the physical values of $f$ and $\qc$. In general,  
by coupling the fields ${\cal{A}}_{\mu}$ and  $\widetilde M_q^{V,A}$ 
 to quark loops, the chiral Lagrangian terms  and their coefficients
within   the light sector   can be obtained.

\begin{figure}[bt]
\begin{center}
  \epsfig{file=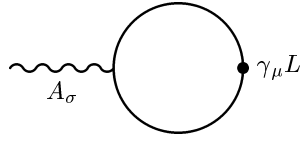,width=4cm}
\caption{Feynman diagram for bosonization of the left-handed current
to order ${\cal O}(p)$.}\label{fig:AL}
\end{center}
\end{figure}

Similarly, we may bosonize the weak currents.
The  left handed current can be written 
\begin{equation}
\bar q_L \gamma^\mu \lambda^n q_L \, = \,  \,
\bar \chi_L  \gamma ^\mu \Lambda^n  \, \chi_L  \; ;
\qquad
\Lambda^n \, \equiv \, \xi^\dagger \lambda^n \,  \xi \; .
\label{leftcur}
\end{equation}
The  lowest order term ${\cal O}(p)$ is obtained when the vertex
$\Lambda^n$
from
(\ref{leftcur}) and axial vertex ($ \sim {\cal A}_\mu$) from
(\ref{chqmR})
are entering a  quark loop (see Fig. \ref{fig:AL}):
\begin{equation}
 j^n_\mu({\cal{A}}) \, = - \, i N_c \int \frac{d^dp}{(2\pi)^d} \,
Tr \, \left[ \left(\gamma_\mu L \, \Lambda^n \right)  S(p) \,
\left( \gamma_\sigma \gamma_5 {\cal{A}}^\sigma \right)  \,  S(p) \,
\right]
  \;  \sim \; Tr \, \left[\Lambda^n \, {\cal{A}}_\mu  \right] \; ,
\label{jXA}
\end{equation}
which coincides with (\ref{Lightcurr}) when (\ref{ASigma})
is used.

As a more non-trivial example,  to obtain a non-zero non-factorizable contribution to
$D^0 \rightarrow K^0 \bar{K^0}$ at tree level,
 one  has to consider the coloured current 
$j^{n,a}_\mu$ to ${\cal O}(p^3)$, involving insertions of the
 ``mass fields'' $\widetilde{M}_q$ in (\ref{cmass}) \cite{EFZ}.
(This coloured current is obtained by Fierz
 transformations of the relevant four quark operator).
 From Fig.~\ref{fig:colcur},
 one obtains the contribution:
\begin{figure}[bt]
\begin{center}
 \epsfig{file=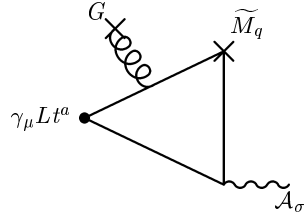,width=4cm}
\caption{Diagram for bosonization of the colour current 
to ${\cal O}(p^3)$.}
\label{fig:colcur}
\end{center}
\end{figure}
\begin{equation}
 j^{n,a}_\mu({\mbox{Fig. 5}}) \,
= i \, \int \frac{d^dp}{(2\pi)^d} \,
Tr \, \left[ \left(\gamma_\mu L \, \Lambda^n t^a \right) \, S(p) \,
\left(\gamma_\sigma \gamma_5 {\cal{A}}^\sigma \right)  \, S(p) \,
 \widetilde{{M}_q} \, S_1(p,G) \right]
  \,  ,
\label{jXG?}
\end{equation}

Summing all six diagrams with permutated vertices compared to the one in
Fig. \ref{fig:colcur} we obtain in total :
\begin{equation}
 j^{n,a}_\mu(G,\xi, {\cal{A}},\widetilde{{\cal{M}}_q}) \,
=  \,  \frac{g_s}{12 m} \frac{1}{16 \pi^2} \, G^{a, \kappa \lambda}
\left[ i \varepsilon_{\mu \rho \kappa \lambda} \;
T_{\varepsilon}^{n,\rho}
\, + \, \left(g_{\mu \kappa} g_{\rho \lambda} \; - \;
g_{\mu \lambda} g_{\rho \kappa}  \right) T_g^{n,\rho} \right] \; ,
\label{jXG}
\end{equation}
where (we have used the analytical
computer program FORM \cite{FORM})
\begin{equation}
T_{\varepsilon}^{n,\rho} \, = \, 4 \, S^K_\rho \, - 3 (S^L_\rho +
S^R_\rho)
\; , \;\;
T_g^{n,\rho} \, =  S^L_\rho \, - \,  S^R_\rho \; .
\label{Teg}
\end{equation}
The $S's$ are chiral Lagrangian terms:
\begin{equation}
\begin{split}
S^L_\rho \, & \equiv \, Tr \left[ \Lambda^n  {\cal{A}}_\rho \,
 \widetilde{M}_q^L \right]
  \,  = \, \frac{1}{2i} Tr \left[ \lambda^n (D_{\rho} \Sigma)
\, {\cal{M}}_{q}^{\dagger}  \right]  \; ,  \\
S^R_\rho \, & \equiv \, Tr \left[ \Lambda^n \, \widetilde{M}_q^R
 \, {\cal{A}}^\rho    \, \right]
  \,  = \, \frac{-1}{2i} Tr \left[ \lambda^n \, {\cal{M}}_{q} \,
(D_{\rho} \Sigma^\dagger ) \,  \right] \; ,  \\
S^K_\rho \, & \equiv \, \frac{1}{2}Tr \left[ \Lambda^n
\left(  {\cal{A}}^\rho \,  \widetilde{M}_q^R
\, + \,  \widetilde{M}_q^L  \, {\cal{A}}^\rho \right)   \,
\right]
  \\
 &= \, \frac{1}{4i} Tr \left[ \lambda^n \left( (D_{\rho} \Sigma)
\, \Sigma^\dagger {\cal{M}}_{q} \Sigma^\dagger \, - \,
 \Sigma {\cal{M}}_{q}^\dagger  \Sigma  (D_{\rho} \Sigma^\dagger)
\right)  \right] \; .
\label{SLRK}
\end{split}
\end{equation}
The current (\ref{jXG}) has to be be combined with the left-handed
colour current for $D$-meson decay later given 
in (\ref{1G}) to obtain a contribution to  
$D^0 \rightarrow K^0 \bar{K^0}$ \cite{EFZ}.

\section{The Heavy -  Light Chiral Quark Model (HL$\chi$QM)}\label{sec:HCQM}

\subsection{The Lagrangian for HL$\chi$QM}

Our starting point is the following Lagrangian containing both quark
 and meson fields:
\begin{equation}
{\cal L} =  {\cal L}_{HQEFT} +  {\cal L}_{\chi QM}  +   {\cal L}_{Int} \; ,
\label{totlag}
\end{equation}
where \cite{neu}
\begin{equation}
{\cal L}_{HQEFT} =  \, \pm \, \overline{Q_{v}^{(\pm)}} \, i v \cdot D \, Q_{v}^{(\pm)} 
 + \frac{1}{2 m_Q}\overline{Q_{v}^{(\pm)}} \, 
\left( - C_M \frac{g_s}{2}\sigma \cdot G
 \, +   \, (i D_\perp)_{\text{eff}}^2  \right) \, Q_{v}^{(\pm)}
 + {\cal O}(m_Q^{- 2})
\label{LHQEFT}
\end{equation}
is the Lagrangian for heavy quark effective field theory (HQEFT).
The heavy quark field  $Q_v^{(+)}$
annihilates  a heavy quark   with velocity $v$ and mass
$m_Q$.  Similarly,   $Q_v^{(-)}$ annihilates  a heavy anti-quark. 
Moreover,  
$D_\mu$ is the covariant derivative containing the gluon field
(eventually also the photon field), and 
$\sigma \cdot G = \sigma^{\mu \nu} G^a_{\mu \nu} t^a$, where 
$\sigma^{\mu \nu}= i [\gamma^\mu, \gamma^\nu]/2$,
and  $G^a_{\mu \nu}$
is the gluonic field tensor.
This chromo-magnetic term has a factor $C_M$, being one at tree level,
 but slightly modified by perturbative QCD.(When the covariant
 derivative also 
contains the photon field, there is also a corresponding magnetic term
$\sim \sigma \cdot F$, where $F^{\mu \nu}$ is the electromagnetic tensor).
 Furthermore,  
$(i D_\perp)_{\text{eff}}^2 =
C_D (i D)^2 - C_K (i v \cdot D)^2 $. At tree level, $C_D = C_K = 1$.
Here, $C_D$ is not modified by perturbative QCD, while $C_K$ is different 
from one due to perturbative QCD corrections for $\mu < m_Q$ \cite{GriFa}.
We observe that soft gluons
coupling to a heavy quark is suppressed by $1/m_Q$, since to leading
order the vertex is proportional to $v_\mu v_\nu G^{a\mu\nu}= 0$,
 $v_\mu$ being the heavy quark velocity.

In the heavy -  light case, the generalization of the
 meson -  quark interactions in the pure light sector  $\chi$QM
is given by the following $SU(3)$ 
invariant Lagrangian:
\begin{equation}
{\cal L}_{Int}  =   
 -   G_H \, \left[ \chibar_k \, \overline{H_{v k}^{(\pm)}} \, Q_{v}^{(\pm)} \,
  +     \overline{Q_{v}^{(\pm)}} \, H_{v k}^{(\pm)} \, \chi_k \right] \,  ,
\label{Int}
\end{equation}
where $k$ is a triplet $SU(3)$- index and $G_H$ is a  coupling constant.
Note that in \cite{HLchiqm}, $G_H=1$ is used. However,
 in that case
one used a renormalization factor  for the heavy meson fields $H_v$,
which is equivalent. 
The interaction  Lagrangian  (\ref{Int}) can, as
for the $\chi$QM, be obtained from a NJL model.
 This has been done in
\cite{HLchiqm} (-as for the light sector \cite{bijnes}).

\subsection{Bosonization within the HL$\chi$QM}\label{sec:strong}
 
The interaction term ${\cal{L}}_{Int}$ in (\ref{Int}) can now 
be used to bosonize the model, i.e. 
integrate out the quark fields. This can be done  
 in terms of Feynman diagrams as we do here, by attaching the
 external fields $H_v^{a}, \overline{H_v^{a}},  {\cal{V}}^\mu,
  {\cal{A}}^\mu$ and $\widetilde{M}_q^{V,A}$ of section II-B to quark
  loops, and  using (\ref{chqmR}) and (\ref{Int}). In this way
one obtains the 
strong chiral Lagrangian (\ref{LS0}) and terms of higher order
in the heavy light sector .
Some of the loop integrals will be divergent and have  
 to be related to physical parameters, as for the pure
 light sector \cite{chiqm,pider,epb,BEF}.
As the pure light sector is a part of our model, we have to keep the 
relations in (\ref{fpidiv}) and (\ref{I1}) from the pure light sector
within the heavy light case studied here.
The $1/m_Q$ terms will not be discussed in this section, but will be considered
later in section  V.

From the diagrams in Fig.~\ref{fig:va}
 we obtain the identification for the kinetic term , which by Ward
 identities is the same as for the term with the vector field ${\cal V}_\mu$ 
attached to the light quark:
\begin{equation}
 -  iG_H^2N_c \, (I_{3/2}  +   2mI_2  +  i \fr{(8-3 \pi)}{384 N_c m^3}\gc) =   1 \; ,
\label{norm}
\end{equation}
where $I_2$ is given in (\ref{I2int}) and 
\bea
I_{3/2}\, &\eq &\, \int\fr{d^dk}{(2\pi)^d}\fr{1}{(v\cdot k)(k^2 -
  m^2)} \; ,
\label{I3}
\eea
which formally depends on $v^2$ which is  equal to one. 
From the same diagram, with the axial field ${\cal A}_\mu$ attached,
 we obtain
 the following identification for the
 axial vector coupling $g_{\cal A}$ :
\begin{figure}[t]
\begin{center}
   \epsfig{file=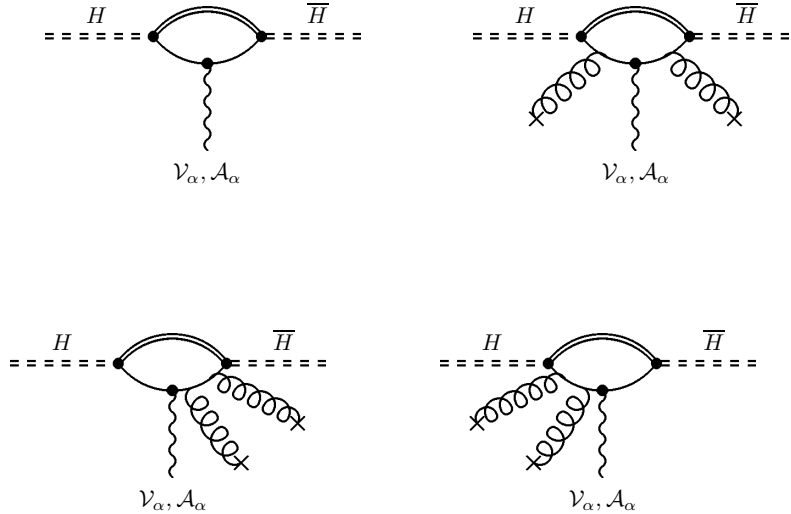,width=10cm}
\caption{Bosonization in the strong sector to obtain eq. (\ref{LS0}).}
\label{fig:va}
\end{center}
\end{figure}
\begin{equation}
 g_{\cal A}\eq i G_H^2 N_c \left( \fr{1}{3}I_{3/2} -   2mI_2  - i  \fr{m}{12
 \pi}
 - i \fr{(8-3 \pi)}{384 N_c m^3} \gc \right) \; ,
\label{ga}
\end{equation}

As $I_1$ and $I_2$ are related to the quark
condensate and $f_\pi$ respectively, the (formally) 
linearly divergent integral  
$I_{3/2}$ is related to
$\delta g_{\cal A} \equiv 1  -  g_{\cal A}$, which is found
 by eliminating $I_2$
from eqs. (\ref{norm}) and (\ref{ga}) :
\begin{equation}
 \delta g_{\cal A} =   - \frac{4}{3} i G_H^2N_c \, \left(I_{3/2} 
 -i \frac{m}{16 \pi} \right)  \; .
\label{dga}
\end{equation}
Note that the gluon condensate drops out here.
Within a primitive cut-off regularization, $I_{3/2}$ is (in the leading
 approximation) proportional to the cut-off in first power:
\bea
I_{3/2}= i \frac{\Lambda}{16 \pi} \left(1 + 
{\cal O}(\frac{m}{\Lambda})\right) \; \, , 
\eea
where the cut-off $\Lambda$,
 is of the same order 
as the chiral symmetry breaking scale $\Lambda_\chi$.
In contrast, $I_{3/2}$ is finite and proportional to $m$
in dimensional regularization.
Note that the cut-off $\Lambda$ is only used in
qualitative considerations here and in subsection IV-D.
Anyway, $I_{3/2}$ is determined by the physical value og $\ga$.

When attaching $\widetilde{M}_q^V$ like in Fig.~\ref{fig:va}
instead of vector or axial vector fields one finds for the mass
correction
term in (\ref{LS0}):
\begin{equation}
2 \lambda_1 \eq  i G_H^2 N_c (I_{3/2} -   2mI_2  - i \frac{m}{8 \pi}
  -   \fr{i (3 \pi -4)}{192 N_c m^3}\gc) \; .
\label{lam1}
\end{equation}
The electromagnetic $\beta$ term in (\ref{LS0}) is obtained
 by considering  diagrams  like those in
figure~\ref{fig:va}, but with the vector and axial vector fields
${\cal V}_{\mu}$ or  ${\cal A}_\mu$ 
 replaced by a photon field tensor:
\begin{equation}\label{betaLOR}
\beta =
\fr{G_H^2}{2}\left\{-4 i N_c I_2 +\fr{N_c}{4\pi}
-\left(\fr{32+3\pi}{576  m^4}\right)\gc\right\}\, .
\end{equation}

\begin{figure}[t]
\begin{center}
   \epsfig{file= 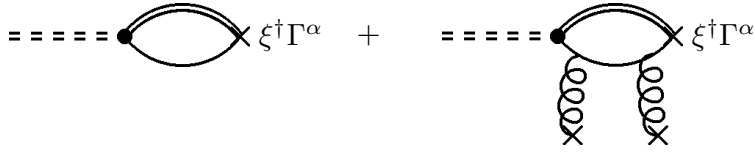,width=10cm}
\caption{Diagrams for bosonization of the left handed quark current.}
\label{fig:f+}
\end{center}
\end{figure}

Within the full theory (SM) at quark level, the weak current is :
\begin{equation}
J_k^\alpha  =  {\overline{q_k}_L}\,\gamma^\alpha\, Q
\label{Lcur}
\end{equation}
where $Q$ is the heavy quark field in the full theory.
Within HQEFT this current will,  below the renormalization scale
 $\mu  =   m_Q \, (=m_b, m_c)$, be modified in the following way:
\begin{equation}
J_k^\alpha
 =  {\chibar}_h \xi^{\dagger}_{hk} \Gamma^\alpha  Q_v     
\, + {\cal O}(m_Q^{-1}) \; ,
\label{modcur}
\end{equation}
The operator in equation (\ref{modcur}) can be bosonized by calculating the 
Feynman diagrams shown in Fig.~\ref{fig:f+}
which gives the bosonized current in (\ref{J(0)}) with:
\begin{equation}
\alpha_H \eq  -  2iG_HN_c\left(  -  I_1  +  mI_{3/2} +
 \fr{i(3 \pi -4)}{384 N_c m^2} \gc\right) \; .
\label{alphaH}
\end{equation}

 To first order in the chiral expansion we obtain the bosonized current
obtained by attaching one extra field  ${\cal A_\nu}$ to the quark
loops in Fig.~\ref{fig:f+}:
\begin{equation} 
J_k^\alpha (1) \, = \, \fr{1}{2}Tr\left[\xi^{\dagger}_{hk}\Gamma^\alpha
H_{vh}(\alpha_{H \gamma}^{(1)}\gamma^\nu\gamma_5 +
\alpha_{H v}^{(1)} v^\nu\gamma_5) {\cal A_\nu} \right] \; , 
\label{cJ1}
\end{equation}
where the quantities $\alpha_H^{(1)}$ are given by expressions 
similar to (\ref{alphaH}).

The coupling $\alpha_H$ in (\ref{J(0)}) is related to  the physical
 decay constants $f_H$  and $f_{H^*},$ in the following way (for $H=B,D$):
\bea
\bra 0| \overline{u}\gamma^\alpha \gamma_5 b|H \ket 
 =   -  2\,\bra 0| J_a^\alpha|H \ket\, 
 =  iM_Hf_H v^\alpha \; , 
\eea
Taking the trace over the gamma matrices in (\ref{J(0)}),
 we obtain a relation for
$\alpha_H$ and the  relations between the heavy meson decay
constants $f_H$ and $f_{H^*}$ (for  $H=B,D$) :
\begin{equation}
\alpha_H =  \fr{f_H\sqrt{M_H}}{C_\gamma (\mu ) + C_v(\mu )} =
\fr{f_{H^*}\sqrt{M_{H^*}}}{C_\gamma(\mu)}\;\;  ,
\label{fb}
\end{equation}
where the model dictates us to put $\mu  =   \Lambda_\chi$.
Later, in section V-B, we will see how chiral corrections and $1/m_Q$ corrections
modify this relation.

\subsection{Constraining the parameters  of the HL$\chi$QM}\label{sec:param}

The gluon condensate can be related to the chromomagnetic
interaction : 
\begin{equation}
\mu_G^2(H) =  \fr{1}{2M_H}C_{\text{M}}(\mu)
\bra H|\bar{Q_v}\fr{1}{2}\sigma\cdot GQ_v|H\ket\, , \label{gc2}
\end{equation} 
where the coefficient $C_{\text{M}}(\mu)$ contains the short distance
 effects down to
the scale $\mu$ and has been calculated to next to leading order ($NLO$) 
\cite{neub}.
The chromomagnetic operator is responsible for the splitting between
the $1^- $ and $0^-$ state, and is  known from spectroscopy, 
\bea
\mu_G^2(H) \,  = \, \frac{3}{2} m_Q (M_{H^*} -  M_H)
 \; .  
\label{mug2exp}
\eea 

An explicit calculation of the matrix element 
 in equation (\ref{gc2}) gives
\begin{equation}
\label{mug2}
\mu_G^2  =   \eta_2 \frac{G_H^2}{m} \gc \; , \quad 
\text{where}\quad\, \eta_2\, \eq\, 
\fr{(\pi +  2)}{32}C_M(\Lambda_\chi)\, .
\end{equation}

Combining  the eqs. (\ref{fpidiv}),
 (\ref{I1}) and  (\ref{mug2}) we obtain the following relations :
\begin{equation}
\label{gcgh}
\gc  =   \fr{\mu_G^2 f^2}{2\eta_2} \, \frac{1}{\rho}  \; \, , \qquad  
G_H^2  =   \fr{2m}{f^2} \, \rho \; \, , 
\end{equation}
where the quantity $\rho$ is of order one and  given by
\begin{equation}
\label{rho}
\rho \, \eq \, \fr{(1 +  3g_{\cal A}) + 
 \frac{\eta_1 \mu_G^2}{\eta_2 m^2}}{4 (1 + \frac{N_c m^2}{8 \pi f^2} )}
\; \, ,
\end{equation}
where
$\eta_1\eq \fr{\pi}{32}$.
In the limit where only the leading logarithmic integral $I_2$ is kept
in (\ref{norm}),
 we  obtain:  
\begin{equation}
g_{\cal A} \rightarrow \, 1 \; , \qquad
\rho \rightarrow \, 1 \; , \qquad \beta \rightarrow \frac{1}{m} \qquad  
G_H \; \rightarrow \; G_H^{(0)} \, \eq \, \fr{\sqrt{2m}}{f} \; \, , 
\label{IRlim}
\end{equation}
which for $g_{\cal A}$ and $\beta$
correspond to the non-relativistic values.

From  the eqs. (\ref{fpidiv}),  (\ref{lam1}), and (\ref{mug2}), we find
\begin{equation}
\label{lamb-ga}
2\lambda_1 = \frac{1}{2}(3 g_{\cal A} -1) + 
\fr{(9 \pi -16)  \mu_G^2}{384\eta_2 m^2} \, .
\end{equation}
 In the limit (\ref{IRlim}) we obtain 
$2 \lambda_1 \rightarrow 1$, as expected. The parameter $\lambda_1$
is related to the mass difference $M_{H_s}- M_{H_d}$.
 To leading order, we obtained the following expression for the $\beta$-term:
\begin{equation}\label{betaLO}
\beta =
\fr{\rho}{m}\left\{1+\fr{N_c m^2}{4\pi f^2}
-\left(\fr{56+3\pi}{576 f^2 m^2}\right)\gc\right\} \; ,
\end{equation}
which is rather sensitive to $m$.
Choosing $m$ in the range 250-300 MeV we find \cite{Beta} 
 $\beta=(2.5 \pm 0.6)$ GeV$^{-1}$ to be compared with
$\beta= (2.7 \pm 0.20)$ GeV$^{-1}$ extracted from experiment.

Using equation (\ref{norm}) and (\ref{I1}) we may write $\alpha_H$ as:
\begin{equation}\label{qcrel}
\alpha_H=\fr{G_H}{2}\left(-\fr{\qc}{m}-2f_\pi^2(1-\fr{1}{\rho})+
\fr{(\pi-2)}{16m^2}\gc\right) \; ,
\end{equation}
Combining (\ref{fb}) with (\ref{qcrel}), we obtain \cite{EFZ} in the
 leading limit (taking into account the logarithmic and quadratic divergent
  integrals only, and 
let $C_\gamma \rightarrow 1$, $C_v \rightarrow 0$ and 
$g_{\cal A} \rightarrow 1 $ as in (\ref{IRlim}) ) we obtain the
 ``Goldberger-Treiman like'' relation: 
\begin{equation}
 f_H\sqrt{M_H}\; \rightarrow \,  -  \frac{\qc}{f_\pi \sqrt{2m}} \;\;  ,
\label{GTrel}
\end{equation}
which gives the  scale for $f_H$ (It is, however, numerically a factor
2 off for the $B$-meson).

Using the relations  (\ref{norm}), (\ref{ga}) and
(\ref{rho}) we obtain for  $\alpha_{H \gamma}^{(1)}$ and $\alpha_{H v}^{(1)}$ 
in (\ref{cJ1}): 
\bea
\alpha_{H \gamma}^{(1)}&&=\fr{2 g_{\cal A}}{G_H}\label{alpha1} \; , \\
\alpha_{H v}^{(1)}&&=\fr{4}{3} G_H 
\left(\fr{f_\pi^2}{2m} \left(\fr{1}{\rho}-1 \right)
+\left[\fr{mN_c}{8\pi}+\fr{(\pi+8)}{256m^3}\gc \right] \right) \; .
\eea
 Moreover, for the mass correction to the weak current given 
in (\ref{J2M}) we find that 
 $\omega_1=-4\lambda_1/G_H$, where $\lambda_1$ is given in equation
(\ref{lam1}) or (\ref{lamb-ga}). The term  $\omega_1^\prime$ 
is subleading in $1/N_c$.

\begin{figure}[t]
\begin{center}
\epsfig{file=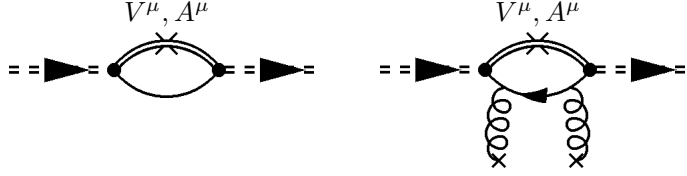,width=10cm}
\caption{Loop diagrams for bosonizing the $b \rightarrow c$ current,
$V^\mu=\gamma^\mu$, $A^\mu=\gamma^\mu\gamma_5$.}
\label{fig:iw}
\end{center}
\end{figure}

The Isgur-Wise function 
 $\zeta (\omega)$  in (\ref{Jbc}) relates all the
form factors describing the processes $B(B^*) \to D(D^*)$ in the
heavy quark limit.
This function can be calculated from the
diagrams shown in Fig.~\ref{fig:iw}. The result  before $1/m_Q$
 and chiral corrections is: 
\begin{equation}
\zeta(\omega)= \fr{2}{1+\omega} \left(1-\rho \right)+
 \rho \, r(\omega)\, , 
\label{iw}
\end{equation}
where $\rho$ is given in (\ref{rho}) and 
$r(\omega)$ is the same function appearing in loop calculations of the
anomalous dimension in HQEFT :
\begin{equation}
r(\omega)=\fr{1}{\sqrt{\omega^2-1}}\, 
\text{ln}\left(\omega+\sqrt{\omega^2-1}\right) \; .
\end{equation}
Note that  $\zeta(1) = 1$ as it should.

\subsection{The formal limit $m \rightarrow 0$}
\label{chiral}

In this subsection we will discuss the limit of restauration of chiral
symmetry, i.e. the limit $m \rightarrow 0$ \cite{Beta}. In order to do this, we
have to consider the various constraints obtained when constructing the
HL$\chi$QM \cite{AHJOE}.

Looking at the equations (\ref{fpidiv}) and (\ref{I1}), one may worry 
that $\qc$ and $f$ behaves like $1/m$ in  the  limit $m \rightarrow
0$ unless one assumes that $\gc$ also go to zero in this limit.
We should stress that the exact limit $m = 0$ cannot be
taken because our loop integrals will then be meaningless. Still we may let
$m$ approach zero without going to this exact limit.
In the pure light sector (at least when vector mesons are not
included)
 there are no restrictions on how $\gc$
might go to zero. In the heavy light sector we have in addition to
(\ref{fpidiv}) and (\ref{I1}) also the relations (\ref{norm}) and 
(\ref{ga}),
 which put restrictions on
the behavior of the gluon condensate $\gc$ for small masses.
As $\gc$ has dimension mass to the fourth power, we find that 
$\qc$ and $f^2$ may
go to zero if $\gc$ goes to zero as $m^4$ or $m^3 \Lambda$ (eventually
combined with $\ln (m/\Lambda)$). However, the behavior $m^3 \Lambda$ is
inconsistent with the additional equations 
(\ref{gcgh}) and  (\ref{rho}).
Still, from all equations (\ref{norm}), 
(\ref{gcgh}), (\ref{rho}), we find the  possible solution
\bea
\gc = \hat{c} \, N_c m^4 K(m) \; \; \text{, where}\quad 
K(m) \equiv (-4i I_2 + \frac{1}{8 \pi}) \; ,
\label{chilim}
\eea
and $\hat{c}$ is some constant.
Then we must have the following behavior for $G_H^2$, $\ga$ and $\mu_G^2$
when $m$ approaches zero:
\bea
G_H^2 \sim \frac{1}{N_c \Lambda} \quad , 
 \; (1+ 3\ga)  \sim \frac{m}{\Lambda} K(m) \;
\quad , \; \; \mu_G^2 \sim  \frac{m^3}{\Lambda} K(m) \; \, ,
\label{chilimR}
\eea
with some restrictions on the proportionality factors.
Here, the regularized $I_2$ is such that for small $m$,  $K(m) = (c_1
+c_2 \ln m/\Lambda)$, $c_1$ and $c_2$ being constants.
 The behavior of  $G_H^2$ is in
 agreement with Nambu-Jona-Lasinio 
models \cite{bijnes}. Note that in our model, $\delta \ga \rightarrow
4/3$ (corresponding to $\ga \rightarrow
-1/3$) for  $m \rightarrow 0$, in contrast to 
$\delta \ga \rightarrow 2/3$  for a free  Dirac
particle with $m=0$.
Note that in \cite{AHJOE} we gave the variation of the gluon
condensate with $m$ for a fixed value of $\mu_G^2$. For the
considerations in this subsection,
 we have to let  $\mu_G^2$  go to zero with $m$
in order to be consistent. When $m \rightarrow 0$, we also find that 
$\beta \rightarrow 1/\Lambda$, provided that the coefficient $\hat{c}$
in (\ref{chilim}) is fixed to a specific value (which is
$\hat{c}= 576/(3 \pi +32) \simeq (1.93)^4$).

\section{$1/m_Q$ corrections within the HL$\chi$QM}\label{sec:mq}

\subsection{Bosonization of the strong sector}

To order $1/m_Q$
one obtains further contributions to chiral Lagrangians (see ref.
\cite{AHJOE} and references therein):
\begin{figure}[t]
\begin{center}
   \epsfig{file=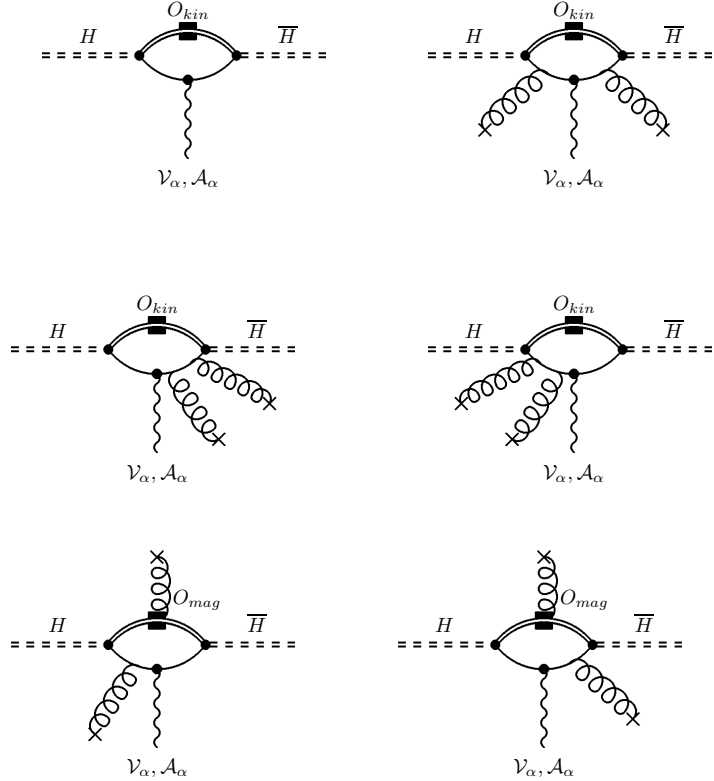,width=9cm}
\caption{Diagrams responsible for $1/m_Q$ terms in the chiral Lagrangian.}
\label{fig:kin}
\end{center}
\end{figure}
\bea
{\cal L}_{Str}\, &&= 
  \, -  \frac{\varepsilon_1}{m_Q} 
Tr\left[\overline{H_{k}}(iv\cdot D)H_{k}\right]\, + \,
  \frac{\varepsilon_1}{m_Q} Tr\left[\overline{H_{k}}H_{h}v_\mu {\cal V}^\mu_{hk}
\right] \nonumber \\
 &&+ \,  \frac{g_1}{m_Q} 
Tr\left[\overline{H_{k}}H_{h}\gamma_\mu\gamma_5 {\cal
A}^\mu_{hk}\right]\nonumber
+\fr{\varepsilon_2}{m_Q}
Tr\left[\overline{H_{k}}\sigma^{\alpha\beta}iv\cdot D H_{k}\sigma_{\alpha\beta}\right]
 \,\nonumber \\&&
 -\, \fr{\varepsilon_2}{m_Q}
Tr\left[\overline{H_{k}}\sigma^{\alpha\beta}v_\mu {\cal V}^\mu_{hk} 
\sigma_{\alpha\beta}H_{h}
\right]+\, \fr{g_{2}}{m_Q}Tr\left[\overline{H_{k}}
\gamma_\mu\gamma_5 {\cal A}^\mu_{hk}H_{h}\right] \, + ....
 \label{LS1}
\eea
where the ellipses indicate other terms (of higher order, say), and 
$D_\mu$ contains the photon field.
The new terms of order $1/m_Q$ in (\ref{LS1}) are a consequence of
 the chromomagnetic interaction $O_{mag}$
(the second term in equation (\ref{LHQEFT})),
 and the kinetic interaction $O_{kin}$ (the third term in (\ref{LHQEFT})).
 Calculating the diagrams of Fig.~\ref{fig:kin} and 
eliminating the divergent integrals  and 
using equations  (\ref{I2int}), (\ref{I1}) and (\ref{dga}),
gives for example
\bea
g_1&=&\,m-G_H^2\left(\fr{\qc}{12m}+\fr{f_\pi^2}{6}+
\fr{N_cm^2(3\pi+4)}{48\pi}- \nonumber
\fr{C_K}{16}(\fr{\qc}{m}+3f_\pi^2)\right.\\&&\left.\qquad\qquad\quad
+\fr{(C_K-2 \pi)}{64 m^2}\gc\right) \;  , \\
g_2&=& \fr{(\pi + 4)}{(\pi+2)} \fr{\mu_G^2}{6 m}
\; .
\eea
As the $1/m_Q$ terms break heavy quark spin symmetry, the chiral Lagrangian
in (\ref{LS1}) will split in $H(0^-)$ and $H^*(1^-)$ terms respectively.

\subsection{The weak current to order $1/m_Q$}\label{sec:mqweak}

In HQEFT the weak vector current at order $1/m_Q$ is \cite{neu}:
\begin{equation}
J^\alpha=\sum_{i=1,2}C_i(\mu)J_i^\alpha+\fr{1}{2m_Q}\sum_jB_j(\mu)O_j^\alpha
+\fr{1}{2m_Q}\sum_kA_k(\mu)T_k^\alpha \; ,
\label{JQ}
\end{equation}
where the first terms are given in  (\ref{Gamma}) and  (\ref{modcur}), 
the $B_j$'s and $A_j$'s are Wilson coefficients, and the $O_j^\alpha$'s
are two quark operators
\bea
& O_1^\alpha=\bar{q}_L\gamma^\alpha i\slash{D}Q_v \; , \quad & 
O_4^\alpha=\bar{q}_L\gamma^\alpha 
(-iv\cdot\overleftarrow{D})Q_v \; , \nonumber\\
& O_2^\alpha=\bar{q}_Lv^\alpha i\slash{D}Q_v\quad \; , 
& O_5^\alpha=\bar{q}_Lv^\alpha 
(-iv\cdot\overleftarrow{D})Q_v\nonumber \; , \\
& O_3^\alpha=\bar{q}_LiD^\alpha Q_v \; , \quad & O_6^\alpha=\bar{q}_L 
(-i\overleftarrow{D}^\alpha)Q_v   \; ,
\label{B}
\eea
The operators $T_k$ are nonlocal and is
a combination of the leading order currents $J_i$ and a term of order
$1/m_Q$ from the effective Lagrangian (\ref{LHQEFT}).

\begin{figure}[t]
\begin{center}
   \epsfig{file=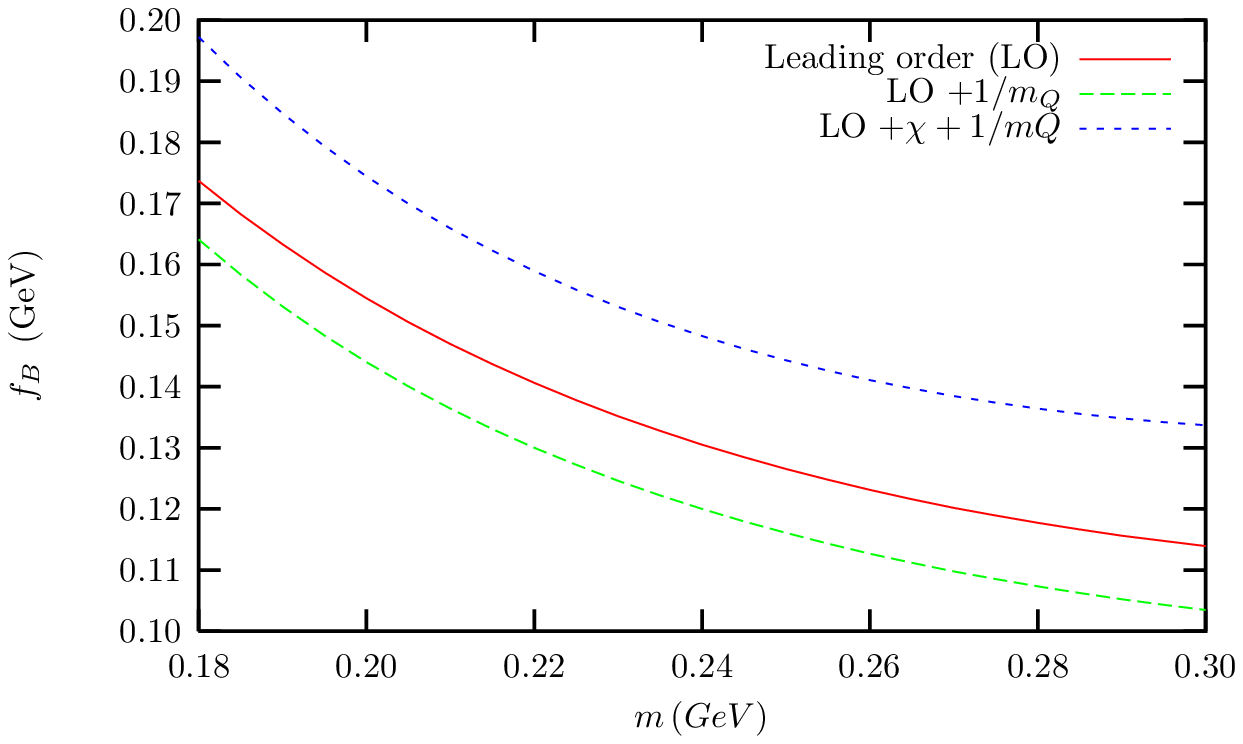}{fig:Fb}
\caption{$f_B$ as a function of $m$ for $\qc^{1/3}$ = -240 MeV.}
\label{fig:Fb}
\end{center}
\end{figure}

Combining  (\ref{JQ})  with (\ref{J(0)}), and adding chiral corrections
and the $1/m_Q$ corrections indicated in
 (\ref{modcur}),  we obtain for $H = B,D$: 
\bea
f_H = \frac{1}{\sqrt{M_H}}\left[\left(C_\gamma (\mu ) + C_v(\mu )\right)  
\alpha_H  + \frac{\eta_Q}{m_Q} + \frac{\eta_\chi}{32\pi^2 f_\pi^2}
 \right]
\; ,
\label{fH}
\eea
where $C_{\gamma, v}$ are defined in (\ref{Gamma}).
Here the model dictates us to put $\mu  =   \Lambda_\chi$.
The quantities $\eta_Q$ and  $\eta_\chi$ are given in \cite{AHJOE}.
One should note that the quantities $\eta_Q$ for $Q=b,c$ depend on the
Wilson coefficients $C_i, B_i, A_i$ in (\ref{JQ}) and some hadronic
parameters,
for instance $\varepsilon_{1,2}$ from (\ref{LS1}). The Wilson
coefficients
entering $f_H$ depends on $m_Q$ through $\ln (m_Q/\mu)$, and therefore
$f_H$ is a complicated function of $m_Q$, $\qc$, $\ga$, $f_\pi$,
and the constituent light quark mass $m$. Note that 
$\gc^{1/4}$ is fixed to be around 320MeV.
 In Fig.~\ref{fig:Fb}, $f_B$ is
 plotted as function of $m$ for standard values of the other
 parameters. One should note that bigger values of $|\qc|$ give higher
 values of $f_B$.
 For a discussion of the numerical values of our
 parameters, see \cite{AHJOE} and \cite{Beta}.

\section{Applications}

In this section we focus on the chiral quark model aspects, especially
contributions proportional to the gluon condensate. There are
always additional 
chiral loop corrections which can be found in
\cite{BEF,AHJOE,ahjoeB,EFH} and references therein.

\subsection{$\bbar$ mixing and heavy quark effective theory}

At quark level, the standard
effective Lagrangian describing $\bbar$ mixing is \cite{EffHam}
\bea
{\cal L}_{eff}^{\Delta B= 2} \, = \,
 - \, \fr{G_F^2}{4\pi^2} M_W^2 \left(V_{tb}^*V_{tq}\right)^2 
\,S_{IL}\left(x_t \right)\,\eta_B \,
b(\mu) \; Q_B \; , 
\label{LB2}
\eea
where  $G_F$ is Fermi's coupling constant, the $V$'s are 
KM factors 
 (for 
which $q=d$ or $s$ for $B_d$ and $B_s$ respectively)
 and $S_{IL}$ is the Inami-Lim function due to
 short distance electroweak
loop effects for the box diagram. 
The quantity $Q_B \equiv Q(\Delta B=2)$ is a four quark
operator 
\begin{equation}
Q_B= \overline{q_L}\,\gamma^\alpha\, b_L \;
\overline{q_L}\,\gamma_\alpha\, b_L \; ,
\label{QB2}
\end{equation} 
where $q_L$ $(b_L)$
is the left-handed
projection of the $q$ $(b)$-quark field.
The quantities $\eta_B= 0.55\pm 0.01\,$ 
and $b(\mu)$ are calculated in perturbative
quantum chromodynamics (QCD). 
At the renormalization scale 
$\mu~=~m_b$ $(\simeq 4.8~\text{GeV})$ one has $b(m_b)\simeq 1.56$ 
in the naive dimension regularization scheme.
The matrix element of the operator $Q_B$ between the meson
 states is parameterized 
by the bag parameter $B_{B_q}$ :
\begin{equation}
\bra B|Q_B|\overline{B}\ket \eq 
\fr{2}{3} f_B^2 M_B^2 B_{B_q}(\mu) \; \, ,
\label{matrQ}
\end{equation}
where by definition, $B_{B_q} =1 $ within  the factorized limit.
In general, the matrix element of  the operator $Q_B$
is dependent on  $\mu$, and thereby $B_{B_q}$
also depends on $\mu$. 
 As for $K-\overline{K}$
mixing, one defines a renormalization scale independent quantity
\begin{equation}
\hat{B}_{B_q} \equiv b(\mu) B_{B_q}(\mu) \; .
\label{Bhat}
\end{equation}

 The $\Delta B=2$ operator in equation
(\ref{QB2})  can for  $\Lambda_\chi < \mu < m_b$ be written
 \cite{gimenez,mannel} :
\bea
Q_B = \,  C_1 \, Q_1 + C_2 \, Q_2
+\fr{1}{m_b} \sum_i \,h_i X_i
\,+{\cal O}(1/m_b^2) \; \, .
\label{HQ}
\eea
The operator $Q_1$ is $Q_B$ for $b$ replaced by $Q_v^{(\pm)}$,
while $Q_2$ is 
generated within perturbative QCD for $\mu < m_b$.  The operators
 $X_i$ are taking care of $1/m_b$ corrections. The quantities 
$C_1, C_2, h_i$ are Wilson coefficients.
 The  operators are given by  
\bea
&Q_1  &=   2 \;  \overline{q_L}\,\gamma^\mu \,\Qp \; \; 
\overline{q_L}\,\gamma_\mu \,\Qm \, \; , 
\label{Q1}\\
&Q_2  &=   2 \; \overline{q_L}\,v^\mu \,\Qp \; \;
 \overline{q_L}\,v_\mu \,\Qm \; \, ,
\label{Q2}\\
&X_1&= 
 2 \;\overline{q_L}\,iD^\mu \,\Qp 
\;\overline{q_L}\,\gamma_\mu \Qm\ \; + ......
\eea
The explicite expressions for the operators $X_i$ are given in \cite{ahjoeB}.
There are also non-local  operators  constructed as time-ordered
 products of  $Q_{1,2}$ and the first order HQEFT Lagrangian in
(\ref{LHQEFT}).
The Wilson coefficients $C_1$ and $C_2$ have been calculated to NLO
 \cite{gimenez} and for $\mu = \Lambda_\chi$ one has
$C_1(\Lambda_\chi)= 1.22$ and $C_2(\Lambda_\chi)= - 0.15$. 
The coefficients $h_i$ have been calculated to leading order (LO)
in \cite{mannel}.

\begin{figure}[t]
\begin{center}
   \epsfig{file=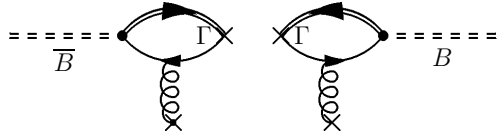,width=7cm}
\caption{Non-factorizable soft gluonic contribution to the
 bag-parameter.
 (Here $\Gamma\eq t^a \,\gamma^\mu\,L$.)}
\label{fig:bbargg}
\end{center}
\end{figure}

In order to find all the matrix element of $Q_{1,2} \,$, one uses the 
following relation between the generators of $SU(3)_c$ ($i,j,l,n$
are colour indices running from 1 to 3):
\begin{equation}
\delta_{i j}\delta_{l n}  =   \fr{1}{N_c} \delta_{i n} \delta_{l j}
 \; +  \; 2 \; t_{i n}^a \; t_{l j}^a \; ,
\label{fierz}
\end{equation}
where $a$ is an index running over the eight gluon charges. This
 means that  by means of a Fierz transformation, the operator $Q_1$
 in (\ref{Q1}) may  also be written in the following way 
(there is a  similar expression for $Q_2$): 
\bea
Q_1^F \; = \; \frac{1}{N_c} Q_1 \; + \;  
  4\, \overline{q_L}\, t^a\, \gamma^\mu  \,\Qp \,
 \overline{q_L} \,t^a\, \gamma_\mu\, \Qm \, .
\label{Q1Fierz}
\eea
The first (naive) step to calculate the matrix element of a four quark 
operator like $Q_1$ is to insert vacuum states between the two currents.
This factorized limit
means to bosonize the two currents in $Q_1$
 and multiply them (see (\ref{modcur})).
 The second operator in (\ref{Q1Fierz}) is genuinely non-factorizable.
In the approximation where only the lowest gluon condensate is
 taken into account, the last term in (\ref{Q1Fierz}) can be written in a 
{\it quasi-factorizable} way by
bosonizating the heavy-light colour current 
with an extra  colour matrix $t^a$ inserted and with an extra gluon 
emitted as shown in Fig.~\ref{fig:bbargg}.

We find  the bosonized colour current:
\bea
\left(\overline{q_L} t^a\,\gamma^\alpha Q_v^{(\pm)}\right)_{1G} 
\;   \longrightarrow \;
 -\fr{G_H g_s}{8}\,G_{\mu\nu}^a 
\;
 Tr\left[\xi^\dagger \gamma^\alpha  L \, H^{(\pm)}
\left(\pm i\,I_2\left\{\sigma^{\mu\nu}, \gamma \cdot v\right\}+ \,
\fr{1}{8\pi} \sigma^{\mu\nu} \right)\right] \; ,
\label{1G}
\eea
where $\{ \, , \, \}$ symbolizes an anti-commutator.
The result for the right part of the diagram with $\bar{B}$ replaced by 
$B$  is obtained by 
changing the sign of $v$ and letting ${P_5^{(+)}}\rightarrow {P_5^{(-)}}$.  
Multiplying the coloured currents, we obtain the non-factorizable
parts of $Q_1$ and $Q_2$ to first order in the gluon condensate
by using eq. (\ref{Gluecond}).

 Now the bag parameter can be extracted and  may be written in the 
form:
\bea
\hat{B}_{B_q}=\fr{3}{4} \, \widetilde{b}
 \left[ 1 + \fr{1}{N_c}(1 - \delta_G^B)
+ \fr{\tau_b}{m_b} + 
\fr{\tau_\chi}{32 \pi^2 f^2} \right] \; \, ,
\label{Bhatform}
\eea
where the parameter $\widetilde{b}$ also involves the Wilson coefficients
$C_{\gamma, v}$ defined in (\ref{Gamma}):
\begin{equation}
\widetilde{b} \; = \; b(m_b) \, 
\left[ \fr{C_1-C_2}{(C_\gamma+C_v)^2}\right]_{\mu=\Lambda_\chi} \;  . 
\end{equation}
The soft gluonic  non-factorizable
 effects are given by
\begin{equation}
\delta_G^B \, = \, 
 \frac{N_c \gc}{32\pi^2 f^2 f_B^2} \, \frac{m}{M_B} \,
 \kappa_B \, \left[ \fr{ C_1}{C_1-C_2} \right]_{\mu=\Lambda_\chi} \; \ ,
\end{equation}
where $\kappa_B$ is a dimensionless hadronic parameter which depends on
$m,f,\mu_G^2$ and $\ga$ and is
of order 2.             
Note that
 we are qualitatively in agreement  
with \cite{Mel}, where also a negative contribution to the
 bag factor from soft gluon  effects  is found.
Numerically, $f$ and $f_B$ are of the same order of magnitude, and 
$\delta_G^B$ is therefore suppressed like $m/M_B$ compared to the corresponding
quantity
\begin{equation}
\delta_G^K \, = \, N_c 
 \frac{\gc}{32\pi^2 f^4} 
\end{equation}
found for $K- \overline{K}$ mixing \cite{BEF}.

\begin{figure}[t]
\begin{center}
   \epsfig{file=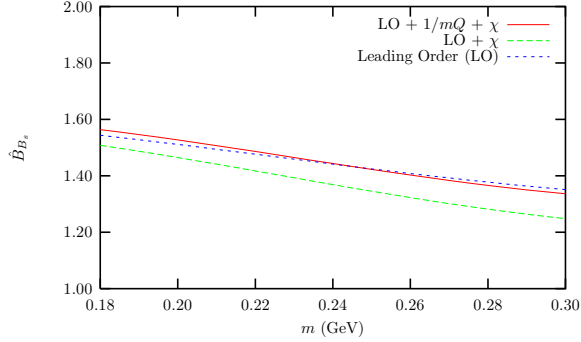,width=8cm}
\caption{The bag parameter $\hat{B}$ for $B_s$ as a function of $m$.}
\label{fig:Bbs}
\end{center}
\end{figure}

 However, one should note that $f_B$
scales as $1/\sqrt{M_B}$ within HQEFT, and therefore 
 $\delta_G^B$ is still formally of order  $(m_b)^0$.
 The quantity $\tau_b$  represents the $1/m_b$ corrections due to
the operators $X_i$. Furthermore, the quantity $\tau_\chi$ 
represents the chiral corrections (including counterterms)
 to the bosonized versions of $Q_{1,2}$ \cite{ahjoeB}.
The bag parameter $\hat{B}$ is plotted as function of $m$
in Fig.~\ref{fig:Bbs} for the case $B_s$.
Our results are numerically in agreement with 
 lattice results \cite{latt}.

\subsection{The processes $B \rightarrow D^{(*)} \overline{D^{(*)}}$}

It has been observed\cite{EFH} that the procseese
 $\bar{B}_{d,s}^0 \to D_{s,d} \bar{D}_{s,d}$,
 $\bar{B}_{d,s}^0 \to D^*_{s,d} \bar{D}_{s,d}$,
 $\bar{B}_{d,s}^0 \to D_{s,d} \bar{D}_{s,d}^*$, and 
 $\bar{B}_{d,s}^0 \to D^*_{s,d} \bar{D}_{s,d}^*$,
have no factorized contribution from the spectator mechanism.
If one or two of the $D$-mesons in the final state are vectors, there
 are relatively small contributions from the annihilation mechanism.
The effective non-leptonic Lagrangian at
quark level has the usual form \cite{EffHam}: 
 \begin{equation}
 {\mathcal L}_{W}=  - 4 \frac{G_F}{\sqrt{2}} V_{cb}V_{cq}^*
 \sum_i  a_i \; \hat{Q}_i (\mu) \; .
 \label{efflagBDD}
\end{equation}

In our case there are  only two numerically relevant
operators  (for $q = d,s$):
\begin{eqnarray}
\hat{Q}_{1}  =  (\overline{q}_L \gamma^\alpha  b_L )  \; 
           ( \overline{c}_L \gamma_\alpha  c_L )  
   \; \; ; \;
\hat{Q}_{2}  =    ( \overline{c}_L \gamma^\alpha  b_L ) \;  
           ( \overline{q}_L \gamma_\alpha  c_L ) \; .
\label{QhatOp}
\end{eqnarray}  
At $\mu= m_b$, one has $a_2 \sim 1$ and  $a_1 \sim 1/10$.

Using (\ref{fierz}), we obtain
the Fierzed version of the operators $\hat{Q}_{1,2}$:
\begin{eqnarray}
\hat{Q}_{1}^F  = \frac{1}{N_c} \hat{Q}_2 +
  2 ( \overline{c}_L \gamma^\alpha t^a b_L ) \;  
           ( \overline{q}_L \gamma_\alpha t^a c_L ) 
   \nonumber \\
\hat{Q}_{2}^F  =  \frac{1}{N_c} \hat{Q}_1 +
 2 (\overline{q}_L \gamma^\alpha t^a b_L )  \; 
           ( \overline{c}_L \gamma_\alpha t^a  c_L )
\label{FierzBDD}
\end{eqnarray}  
 The genuine non-factorizable $1/N_c$ chiral Lagrangian terms  from
``coloured quark operators''
 can be estimated within the  $HL \chi QM$. However, in order to do
 this we have to treat the effective weak non-leptonic Lagrangian
in (\ref{efflagBDD}) within heavy quark effective theory
(HQEFT)~\cite{neu}.
Then  $b$, $c$, and
$\overline{c}$
quarks are replaced by their corresponding operators  in HQEFT:
\begin{eqnarray}
 b \rightarrow Q_{v_b}^{(+)} \quad , \; \,
 c \rightarrow Q_{v_c}^{(+)} \quad , \; \, 
 \overline{c} \rightarrow Q_{\bar{v}}^{(-)}
\label{HeavyQ}
\end{eqnarray}
up to $1/m_b$ and $1/m_c$ corrections. Then
the effective weak non-leptonic Lagrangian  (\ref{efflagBDD})
 can be evolved down to the scale
 $\mu \sim \Lambda_\chi \sim$1 GeV \cite{GKMWF}.
 At  $\mu=1$ GeV we
have $a_2 \simeq 1.29+0.08 i$, and  $a_1 \simeq -0.35-0.07 i$.
Note that $a_{1,2}$ are complex for $\Lambda_\chi < \mu < m_c$\cite{GKMWF}.

\begin{figure}[t]
\begin{center}
   \epsfig{file=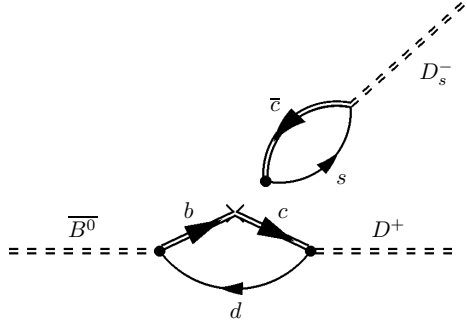,width=7cm}
\caption{Factorized contribution for 
$\overline{B^0}  \rightarrow D^+ D_s^-$
through the spectator mechanism, which does not exist for
the decay mode $\overline{B^0} \rightarrow D_s^+ D_s^-$.
 The double dashed lines represent heavy mesons, the double lines
 represent heavy quarks, and the single lines light quarks.}
\label{fig:bdd_fact}
\end{center}
\end{figure}

The bosonized  factorized weak Lagrangian
corresponding to Fig. \ref{fig:bdd_fact}
 and the operator $\hat{Q}_2$ (with the dominating
 Wilson coefficient $a_2$) is obtained from 
 (\ref{J(0)}),  and (\ref{Jbc}), and  (\ref{efflagBDD}):
 \bea
{\mathcal L}_{W-Fact}^{Bos}(Q_2)= 4 \frac{G_F}{\sqrt{2}} V_{cb}V_{cq}^*
 (a_2+\frac{a_1}{N_c}) \, \zeta(\omega)\frac{\alpha_H}{2} 
 Tr\left[ \overline{H_c^{(+)}} \gamma^\alpha L  H_{b}^{(+)} \right]
\cdot  Tr\left[\xi^\dagger \gamma^\alpha L 
 H_{\bar{c}}^{(-)} \right] \; ,
\label{factQ2} 
\eea
where $\omega \equiv v_b \cdot v_c= v_b \cdot \bar{v} = M_B/(2M_D)$. 
This   Lagrangian (corresponding to the spectator mechanism)
contributes to the factorized amplitude
for the process $\overline{B^0} \rightarrow D^+ D_s^-$, and is the
starting point for chiral loop contributions of order
$(m_K \ga/4\pi f)^2$ (which are $1/N_c$ suppresed) to the processes
$B \rightarrow D^{(*)} \overline{D^{(*)}}$.

\begin{figure}[t]
\begin{center}
   \epsfig{file=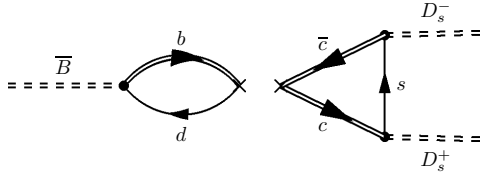,width=7cm}
\caption{Factorized contribution for 
$\overline{B^0}  \rightarrow D_s^+ D_s^-$
through the annihilation  mechanism, which give zero contributions if
both $D_s^+$ and $D_s^-$ are pseudo-scalars.}
 \label{fig:bdd_fact2}
\end{center}
\end{figure}

The bosonized  factorized weak Lagrangian
corresponding to Fig.~\ref{fig:bdd_fact2}
 and the non-dominating
 Wilson coefficient $a_1$ is
\bea
{\mathcal L}_{W-Fact}^{Bos}(Q_1)= 4 \frac{G_F}{\sqrt{2}} V_{cb}V_{cq}^* 
 (a_1+\frac{a_2}{N_c}) \, \zeta(-\lambda)\frac{\alpha_H}{2} 
Tr\left[\xi^{\dagger} \gamma^\mu
L \,  H_{b}^{(+)} \right] \cdot
 Tr\left[
 \overline{H_c^{(+)}} \gamma^\alpha L  H_{\bar{c}}^{(-)} \right]
\eea
where  $\lambda \equiv \bar{v} \cdot v_c = (M_B^2/(2 M_D^2)-1)$.
Unless one or both of the $D$-mesons in the final state 
are vector mesons, this matrix
element is zero due to current conservation, which is
analogous to  the decay mode
$\overline{D^0}  \rightarrow K^0 \overline{K^0}$ \cite{EFZ}.

\begin{figure}[t]
\begin{center}
   \epsfig{file=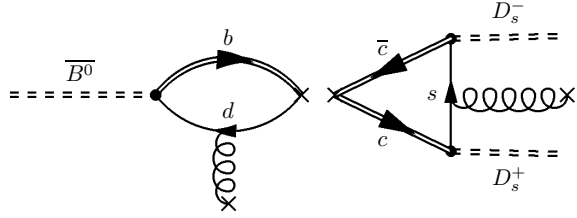,width=8cm}
\caption{Non-factorizable contribution for 
$\overline{B^0}  \rightarrow D_s^+ D_s^-$
through the annihilation mechanism with additional soft gluon emission.
 The wavy lines represent soft
gluons ending in vacuum to make gluon condensates.}
\label{fig:bdd_nfact2}
\end{center}
\end{figure}

The genuine non-factorizable part for 
 $\overline{B^0} \rightarrow D_s^+ D_s^-$
at quark level 
can, by means of Fierz transformations and the identity 
 (\ref{fierz}),   be written 
in terms of  colour currents.
 The  left part in Fig.~\ref{fig:bdd_nfact2} with gluon emission
 gives us the bosonized colour current which is the same as for
 $B-\overline{B}$ mixing in eq. (\ref{1G}). 
For the creation of a $D \bar{D}$ pair in the right part of
 Fig.~\ref{fig:bdd_nfact2}, there is an  analogue of (\ref{1G}), which can
be written:
\bea
\left(\overline{Q_{v_c}^{(+)}} \,t^a \; 
\gamma^\alpha \, L Q_{\bar{v}}^{(-)}\right)_{1G} \;
\;   \longrightarrow \; 
\frac{G_H^2 \, g_s}{32 \pi m}\,G_{\mu\nu}^a
\;Tr\left[ \overline{H_c^{(+)}} \; \gamma^\alpha \, L \,  
  H_{\bar{c}}^{(-)} X^{\mu \nu}  \right]
\label{colcurDD}
\eea

where 
\bea
X^{\mu \nu}  \equiv  \frac{r(-\lambda)}{\pi} \, \sigma^{\mu\nu} \, + \,
\frac{1}{4 (\lambda-1)} \left\{ \sigma^{\mu\nu}, \gamma \cdot t \right\}  
\nonumber
\eea
and $\, t \equiv v_c-\bar{v} \,$ .
Multiplying the currents 
and using (\ref{Gluecond})
we obtain a bosonized effective Lagrangian 
 as the product of two traces. 
Note that our non-factorizable amplitudes 
(proportional to the gluon condensate)  are proportional to the
numerically favourable Wilson coefficient $a_2$.

The gluon condensate contribution obtained from (\ref{1G}) and (\ref{colcurDD})
is a linear combination of  terms like:
 \bea
Tr\left[\xi^{\dagger} \sigma^{\mu \alpha}
L \,  H_{b}^{(+)} \right]  \cdot 
 Tr\left[
 \overline{H_c^{(+)}} \gamma_\alpha L  H_{\bar{c}}^{(-)} \gamma_\mu
 \right] \; , \nonumber
\eea
\bea
Tr\left[\xi^{\dagger} \gamma^\mu L \,  H_{b}^{(+)} \right] \cdot
 Tr\left[ \overline{H_c^{(+)}} \gamma^\alpha L  H_{\bar{c}}^{(-)}
 \sigma_{\mu \alpha} R \right]
\; , \nonumber
\eea
\bea
Tr\left[\xi^{\dagger} \gamma^\mu
L \,  H_{b}^{(+)} \right] \cdot
 Tr\left[
 \overline{H_c^{(+)}} \gamma_5 L  H_{\bar{c}}^{(-)} \gamma_\mu \right]
\; ,  \nonumber
\eea
\bea
Tr\left[\xi^{\dagger}
L \,  H_{b}^{(+)} \right] \cdot
 Tr\left[
 \overline{H_c^{(+)}} \gamma^\alpha L  H_{\bar{c}}^{(-)} \gamma_\alpha
 \right]
\; ,  \nonumber
\eea
\bea
Tr\left[\xi^{\dagger} \sigma^{\mu \alpha}
L \,  H_{b}^{(+)} \right] \cdot
 Tr\left[
 \overline{H_c^{(+)}} \gamma_\alpha L  H_{\bar{c}}^{(-)} \right] 
(\bar{v}-v_c)_\mu
\; ,  \nonumber
\eea
\bea
\varepsilon^{\mu \nu \alpha \lambda} (v_c+\bar{v})_\nu 
Tr\left[\xi^{\dagger} \gamma^\mu
L \,  H_{b}^{(+)} \right] \cdot
 Tr\left[
 \overline{H_c^{(+)}} \gamma^\alpha L  H_{\bar{c}}^{(-)} \gamma_\lambda \right]
\; .
\eea
These terms might have been written down based on the heavy quark
 symmetry, but the HL$\chi$QM selects a certain linear combination to
 be realized.

Our amplitudes for $B \to D \bar{D}$, in terms of  chiral loop and
 gluon condensate contributions, are
sensitive to $1/m_c$ corrections  and  counterterms 
 which are not yet calculated \cite{EFH}.
Operators suppresssed by $1/m_Q$ are obtained by the replacements
\bea
 Q_v^{(\pm)} \to \frac{1}{m_Q} i \gamma \cdot D_\perp Q_v^{(\pm)}
\quad ; \qquad D_\perp^\nu = D^\nu - v^\nu (v\cdot D) \; ,
\label{mQ}
\eea
for one of the heavy quarks 
in  (\ref{QhatOp}) and  (\ref{FierzBDD}). Some new quark operators of
order $1/m_Q$ might also be generated by pQCD for  $\mu < m_Q$.
Counterterms correspond to
 mass insertion of $\widetilde{M}_q \,$, given by (\ref{masst}) and
 (\ref{cmass}), at light quark lines 
in the  diagrams for $B \to D \bar{D}$ this subsection.

\subsection{Other applications}

\begin{figure}[t]
\begin{center}
\epsfig{file=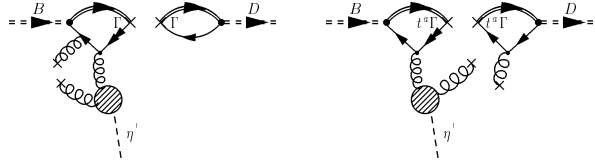,width=8cm}
\caption{Gluon condensate contributions to $B \rightarrow D \eta'$.}
\label{fig:BDeta}
\end{center}
\end{figure}

Within the HL$\chi$QM, the process $B \rightarrow D \eta'$
has been estimated \cite{EHP}.
This is done in two steps. First we calculate the subprocess
$B \rightarrow D g g^*$. 
 Then the virtual gluon  $g^*$ is
attached to the  $\eta' g g^*$-vertex, and the other end in vacuum and
make a gluon condensate together 
with one of the other soft  gluons ($g$) from the
$\eta' g g^*$-vertex. 
Using Fierz transformations for the  four quark operators
for $b \rightarrow c d \bar{u}$, we obtain contributions corresponding
to  Fig.~\ref{fig:BDeta}. 

 We have used existing parameterizations 
of the $\eta' g g^*$-vertex form factor
and   assumed that the  current for $B\rightarrow g^*$ is related 
to the better known case $B\rightarrow \rho$.
It turns out that the ``factorizable'' diagram to the left in 
figure~\ref{fig:BDeta} can be neglected compared to the 
non-factorizable diagram to the right. 
For $m$ in the range 230-270 MeV, we obtained the result \cite{EHP}
$Br(B\rightarrow D\eta^{'}) \,=\, (2.2\, \pm\, 0.4)\,
\times \, 10^{-4} \,$. 
 Here $1/m_Q$ and chiral corrections
are not included.

Heavy to light non-leptonic processes  like
 $B,D \rightarrow K \pi$ cannot in general be treated within the HL$\chi$QM in
 its present form. (See, however, section VII-B). Still,
 semileptonic heavy to light processes
 might be treated at the ``no recoil point'' \cite{AHJOE}.
The  form factors  $f_+(q^2)$ and $f_-(q^2)$ are defined as:
\begin{equation}
\bra \pi^+(p_\pi )|\overline{u} \gamma^\alpha  b|H \ket 
\, = \,  2\,\bra \pi^+(p_\pi )|J_f^{\alpha}|H \ket
\, = \, f_+(q^2)(p_H+p_\pi )^\alpha \, + \, f_-(q^2)(p_H-p_\pi )^\alpha 
\end{equation}
where $p_H^\alpha=M_H v^\alpha$ and the index $a$ corresponds to 
quark flavour $u$ and $q^\mu=p_H^\mu-k_\pi^\mu$.
The form factors get contributions from
$J_f^\alpha (0)$ in (\ref{J(0)}) and $J_f^\alpha (1)$ in (\ref{cJ1})
close to the ``no recoil point'' where $v \cdot p_\pi$ is small: 
\bea
f_+(q^2) \, + \,  f_-(q^2) &=& \fr{-1}{\sqrt{2M_H}f_\pi}\left(C_\gamma +
C_v-g_{\cal A}C_\gamma \right) \alpha_H \; ,
\label{f++f-a}\\
f_+(q^2) \, -  \, f_-(q^2) &=& -C_\gamma\fr{\sqrt{M_H}}{\sqrt{2}f_\pi}
\left(\,\fr{g_{\cal A}\alpha_H}{v\cdot p_\pi} + \alpha_{H \gamma}^{(1)}\,
\right) \; ,
\label{f+-f-b}
\eea
where we have neglected terms of first order in $v \cdot p_\pi$ (where 
$\alpha_{H v}^{(1)}$ contributes). 
 The $1/v\cdot p_\pi$ term in (\ref{f+-f-b})
is due to the $H^*$ pole.
From equation (\ref{f++f-a}) and (\ref{f+-f-b}) we see that
\bea
(f_+(q^2)+f_-(q^2))/(f_+(q^2)-f_-(q^2))\sim 1/M_H
\eea
which is the well known Isgur-Wise scaling
law \cite{isgur2}. The equations for the two form factors $f_+$ and
$f_-$ should be studied further, and chiral corrections and $1/m_Q$
corrections should be added.

\section{Further possible extensions of chiral quark models}
In this section we consider two possible extensions of chiral quark
models which are not yet worked out in detail. The descriptions are
therefore sketchy. 

\subsection{Inclusion of light vectors}
One might  include vectors in the chiral perturbation theory \cite{EckDA}
 and thus it should be possible to use the chiral quark model also in
 this case. We suggest a Lagrangian
\begin{equation}
{\cal L} =  {\cal L}_{mass}+  {\cal L}_{\chi QM}  +   {\cal L}_{IVA} \; ,
\label{totlagV}
\end{equation}
where the interaction between quarks and the vectors and  axial vectors
is given by
\begin{equation}
{\cal L}_{IVA} =  
\chibar \left[h_V \gamma^\mu   V_{\mu}  +  
h_A \gamma_5  A_{\mu} \right]\chi \; .    
\label{chqmV}
\end{equation}
Here $V$ are given as $\Pi$ in (\ref{sigma})
 with  $\pi$ replaced by $\rho$
etc, and 
similarly for the axial vector $A$ where $\pi$ is replaced by $a_1$.
The (bare) mass term is 
\begin{equation}
{\cal L}_{mass} = \bar{m}_V^2 Tr \left[ V_\mu V^\mu \right]
+ \bar{m}_A^2 Tr \left[ A_\mu A^\mu \right] \; .
\label{chqVm}
\end{equation}
After quarks are integrated out, the masses are modified and
identified with the physical ones. Then a kinetic term is also generated:
\begin{equation}
{\cal L}_{kin} = - \frac{1}{2} \left[ V_{\mu \nu} V^{\mu \nu} \right]
\,  -  \, \frac{1}{2} \left[ A_{\mu \nu} A^{\mu \nu} \right] \; ;  
\label{KinVA}
\end{equation}
where for $X=V,A$: 
\begin{equation}
 X_{\mu \nu}  \, = \, \nabla_\mu  X_\nu \, - \, \nabla_\nu  X_\mu \; ,
\label{VTdef}
\end{equation}
and similar for the axial vector.
Here $\nabla$ is a covariant derivative including the goldstones:
\begin{equation}
  \nabla_\mu  X_\nu \, \equiv \partial_\mu X_\nu \, + \, i \,  [{\cal V}_\mu, X_\nu]
 \; .
\label{nabla}
\end{equation}

For the left-handed  current for $vac
\rightarrow X=V,A$ we find the $SU(3)$ octet current
\begin{equation}
J_\mu^n(vac \rightarrow X) = \frac{1}{2} k_X Tr \left[ \Lambda^n X_\mu
  \right] \; ,
\label{CurA}
\end{equation}
 where the quantity $\Lambda^n$ is given by (\ref{leftcur}).

As previuosly, bosonization gives constraints on the parameters of the
vectorial sector.
From normalization of the kinetic term(s) we obtain:
\bea
\frac{f^2 h_V^2}{3 m^2} \left[ 1 \, - \, 
\fr{1}{15 m^2 f^2} \gc \right] \, = \, 1 \; , 
\eea
where $h_A = h_V$ before chiral corrections.
For the currents we obtain
\bea
k_V \, = \, \frac{1}{2} h_V \left(- \fr{\qc}{m}+ f^2 - 
\fr{1}{8 m^2} \gc \right] \; ,
\eea
and similarly , for the axial case:
\bea
k_A \, = \, \frac{1}{2} h_A \left(- \fr{\qc}{m} -3  f^2  + 
\fr{1}{8 m^2} \gc \right] \; .
\eea

The formalism suggested in this subsection might for instance, when combined 
with HL$\chi$QM,  give a
reasonable description of the weak current for $D$-meson decays  $D \rightarrow V$
\cite{Svjet}. This might also  be the case for
 processes like  $D \rightarrow V P$, where
$V$ is a vector meson and $P$ is a pseudoscalar. 
In the last case, non-factorizable contributions can be calculated in
terms of chiral loops and gluon condensates. However, one should keep
in mind that a limitation in this case is that $V$ is (also formally) light
compared to $D$.

\subsection{Heavy to light transitions}

As emphasized in the introduction, the HL$\chi$QM is not suited to
describe $B \rightarrow \pi$ transitions except for semileptonic transitions
close to the no recoil point.
It might therefore be surprising that we consider 
a formalism for  chiral perturbation theory for  $B
\rightarrow \pi$ transitions (and more general  $B
\rightarrow P$ for $P=\pi,K,\eta$), because the involved pion is hard.
 However, in general, in a transition $B \rightarrow $ pions,  
we might have a configuration where one pion is hard and one (or more)
is soft. For such cases we split
 the pseudoscalar sector in hard and
soft pseudoscalars. The soft pseudoscalars are represented as
before , while the hard pseudoscalars are represented by an octet 3$\times$3
matrix $M$ given as $\Pi$ in eq. (\ref{sigma}), but 
transforming as $\Sigma$ under $SU(3)_L \times SU(3)_R$.

Starting with a $\gamma_5$ coupling for quarks coupling to pseudoscalars, we
represent the hard light quark with a quark field $q_n$
\cite{LEET,SCET} and the soft light quarks by the flavour rotated
fields $\chi$ in section III. 
Then we arrive at
an interaction Lagrangian
\bea
{\cal L}_{n} = 
  G_M  \bar{\chi} \left[ \xi^\dagger M R  -  \xi M^\dagger L \right] q_n \; , 
\label{LEET1}
\eea
for a hard light quark $q_n$ entering a hard pion (kaon) with
momentum
$p_M = E \, n$ where  $n$ is a lightlike vector and $E$ is
the energy of the hard pion(kaon). The hard
quark has then momentum $p_q = E \, n + k$, where  $k$ is
of order $\Lambda_\chi \sim$ 1 GeV or smaller.
$G_M$ is a coupling  which has to be determined by some
physical requirements.
 For an outgoing hard quark we have
\bea
{\cal L}_{\bar{n}} = 
  G_M  \bar{q_n} \left[ M \xi^\dagger  R  -  M^\dagger \xi L \right] \chi \; . 
\label{LEET2}
\eea

Now one might
combine (\ref{LEET1}) and (\ref{LEET2}) 
with HL$\chi$QM and use some version of a large energy
effective field theory (LEET) \cite{LEET} to describe the light hard quarks.
 Using  the LEET propagator
\bea 
\frac{\gamma \cdot n}{2 n \cdot k}
\label{LEETPr}
\eea
 for the light hard quark,
we can write down a quark loop diagram for $B \rightarrow P$ 
with a corresponding  amplitude
for the heavy-light weak current  (to leading order) 
\begin{equation}
J_k^\alpha (B \rightarrow P)  \;  =  \; K \,  Tr\left[\Gamma^\alpha 
  H_{v h}\gamma \cdot n \, \xi M^\dagger \right] 
  \; .
\label{JL}
\end{equation}
 Given the
 transformation properties in (\ref{Sitransf}), (\ref{xiHtransf}),
 and (\ref{chHtransf}),
 the current (\ref{JL})   transforms as in (\ref{Jtransf}).

The behaviour of the quantity (form factor) $K$ is known  from
theoretical considerations within 
LEET\cite{LEET} and  soft collinear
 effective theory (SCET)\cite{SCET}:
\bea
K \; \sim \;  E \, \zeta^{(v)}(M_B,E)  \; ,
\label{Kzeta}
\eea
where $\zeta^{(v)}$ is expected to scale as
\bea
\zeta^{(v)}(M_B,E) \; \sim \; \frac{\sqrt{M_B}}{E^2} \; .
\label{zeta}
\eea
Note that a factor $\sqrt{M_B}$  is associated with the heavy ($B$)
meson wave function and similarly  a factor $\sqrt{E}$ with the wave function of
the hard pseudoscalar meson.
Within our framework, $K$ will contain the product of the couplings
$G_H$ and $G_M$, and
 some loop integrals involving the heavy quark propagator, the
 ordinary Dirac propagator for the soft quark, and the LEET propagator
in (\ref{LEETPr}). However, it has  been pointed out that the LEET
propagator is too singular to give meaningful loop integrals\cite{Uglea},
and that the LEET is incomplete\cite{SCET}. Therefore the simple
expression in (\ref{LEETPr}) has to be modified in some way, by
keeping $n^\mu n_\mu=\delta^2 \neq 0$ with $\delta \sim 1/E$, 
by adding a small quantity in the LEET propagator denominator, or by modifying the
 formalism  in other ways. And this modification has to be done such
 that one does not come in conflict with the known  scaling properties of $\zeta^{(v)}$.
Keeping $\delta \neq 0$ and $\delta \sim 1/E$, some of the involved
 loop integrals have the same mathematical form as  those involved
 in the Isgur-Wise
 function in (\ref{iw}), but  with $\omega \to 1/\delta$.  The most
 plausible scenario is that  $G_M \sim E^{-3/2}$. Anyway,
 knowledge of $\zeta^{(v)}$ will put restrictions on $G_M$.

The  $W \rightarrow \pi$ transition is in \cite{BBNS,SCET} 
represented by an integral over a
momentum distribution proportional to $x(1-x)$ dominated at $x \sim
1/2$. 
However, there are also suppressed contributions (for $E >> \Lambda_\chi$)
 from momentum configurations
where one quark (anti-quark)  is hard and the anti-quark (quark)
is soft.
The left-handed current is in this case given by 
\bea
j^l_\mu \; = \bar{q}_n \, \gamma_\mu L \, (\lambda^l \xi) \, \chi
\quad ,\quad  \text{or} \; \qquad
j^l_\mu \; = \bar{\chi} \, (\xi^\dagger \lambda^l) \, \gamma_\mu L \, q_n \; ,
\eea
where $l$ is an $SU(3)$ octet index. These quark currents 
 will, when combined with (\ref{LEET1}) and  (\ref{LEET2}),
generate a bosonized current
\begin{equation}
\Delta J^l_\mu  \,  = \; N \, \tilde{n}^\mu  
Tr\left[ \lambda^l (\Sigma  M^\dagger + M \Sigma^\dagger) \right]
  \; ,\label{Jp}
\end{equation}
where $\tilde{n}$ is another (almost) lightlike
vector with opposite three momentum compared to $n$ such that
$\tilde{n}^2 = \delta^2$ and $\tilde{n} \cdot n = 2 -\delta^2$.
In the most plausible scenario mentioned above $N$ scales as
a constant  ($E^0$),
which is suppressed by $1/E$ compared with the leading order current
 $\sim E \, f^{(0)}_P \, \tilde{n}^\mu$.
The physical decay constant $f_P$ (for $P=\pi,K,\eta$) is within this scheme given
 by   $f^{(0)}_P$ plus the suppressed contribution $\sim N/E$ from (\ref{Jp}).

Now, the product of the currents in (\ref{JL}) and (\ref{Jp}) will
give a factorized $1/E$ suppressed contribution to  $B \rightarrow K \pi$
corresponding to the diagram in Fig.~\ref{fig:bdd_fact}, with $\bar{c}$ and
$c$ replaced by energetic (anti) quarks, $D$ by $\pi$ and $D_s$ by $K$.
This contribution can of course not be distinguished from the standard
factorized contribution. However,
pulling out soft pseudoscalars from $\xi$ and $\Sigma$ in the currents  
(\ref{JL}) and (\ref{Jp}), we obtain 
   $1/E$ suppressed non-factorizable chiral loop
contributions to $B \rightarrow K \pi$. 
Similarly there will be $1/E$
suppressed gluon condensate contributions. Such suppressed terms are not
 in conflict with   QCD factorization~\cite{BBNS}.

\section{Conclusion}

We have presented the main features of chiral quark models, both in
the pure light and the heavy-light sector. Especially, the HL$\chi$QM
seem to work well. In that case,
 it is possible to systematically calculate 
the $1/m_Q$ corrections as well as chiral corrections.
The model may be used to give predictions for many quantities. Especially, 
it is  suitable for calculation of the $B$-parameter for $B-\bar{B}$
mixing \cite{AHJOE}, and for a study of processes of the type $B
\rightarrow D \bar{D}$. 
For heavy to light transitions 
($B \rightarrow K \pi$, say) the HL$\chi$QM 
 cannot be
used in its present form.
It remains to be seen if the 
 extension indicated in sect VII-B
  to incorporate light energetic
quarks will lead to some  understanding of such decays.

In our version of the chiral quark models(pure light and heavy light cases)
soft gluon effects are truncated to include only the secord order
gluon condensate. It has worked reasonably well up to now, 
but one may wonder if this is enogh to accomodate all
effects, for instance when light vectors are included.  Maybe for
instance higher
order gluon condensates could be included, but our simple model will
of course then be much more complicated.

\begin{acknowledgments}

J.O.E.  is supported in part by the Norwegian
 research council
 and  by the European Union RTN
network, Contract No. HPRN-CT-2002-00311  (EURIDICE). He thanks S.
 Bertolini, M. Fabbrichesi, S. Fajfer, I. Picek and A. Polosa for
 collaboration and discussions on chiral quark models.

\end{acknowledgments}

\bibliographystyle{unsrt}

\end{document}